\documentclass[aps,prd,reprint,superscriptaddress,floatfix]{revtex4-2}

\usepackage[english]{babel}
\usepackage{appendix}
\usepackage{graphicx}
\usepackage{dcolumn}
\usepackage{bm}
\usepackage{ytableau}
\usepackage{xcolor}
\usepackage{subcaption}
\usepackage{orcidlink}
\usepackage{amsmath} 
\usepackage{microtype}
\usepackage{amsfonts}
\usepackage{hyperref}
\allowdisplaybreaks

\ytableausetup{mathmode, boxsize=0.2cm}

\begin{document}

\title{Dynamics of spinning test particles in the innermost stable circular orbit \\ of an approximate Kerr-like spacetime with mass quadrupole}

\author{Cristian Herrera-Chaves\,\orcidlink{0009-0001-1127-1880}}\email{cristian.herrerachaves@ucr.ac.cr}
\affiliation{School of Physics, University of Costa Rica}
\author{Francisco Frutos-Alfaro\,\orcidlink{0000-0001-8875-5480}}\email{francisco.frutos@ucr.ac.cr}
\affiliation{School of Physics and Space Research Center, University of Costa Rica \\San José, Costa Rica}
\date{\today}

\begin{abstract}
We study the motion of spinning test particles in a deformed, rotating spacetime described by a metric perturbed up to second order in the mass quadrupole moment. We examine how the parameters of the innermost stable circular orbit (ISCO) ---namely the radius, energy, and angular momentum--- are modified by the particle's spin, contrasting our findings with both the spinless case and standard Kerr benchmarks. By employing effective potential techniques, we consider both co-rotating and counter-rotating configurations, corresponding to parallel and antiparallel alignments of the angular momenta. Using the Euler-Lagrange formalism and Hamiltonian dynamics within the Mathisson–Papapetrou–Dixon framework, we derive an explicit general ISCO equation incorporating the spin and quadrupole parameters, alongside analytical expressions for the conserved orbital quantities. Furthermore, by taking the Schwarzschild, Lense-Thirring and Kerr limits within our generalized framework, we establish new ISCO equations and compact expressions for the energy and angular momentum that extend these fundamental spacetimes to the case of spinning test particles.
\end{abstract}

\maketitle

\section{Introduction}\label{sec:intro}

The spacetime surrounding an astrophysical compact object contains information about its internal structure through a hierarchy of relativistic multipole moments. In general relativity, stationary and axisymmetric spacetimes are characterized by their Geroch–Hansen multipole moments, which determine the gravitational field outside the source \cite{geroch,hansen}. It is widely acknowledged nowadays that the Kerr metric \cite{kerr} does not accurately describe the spacetime of a realistic rotating astrophysical object \cite{hernandez}. The relationship between its multipole moments and its angular momentum may fail to characterize the structure of the external gravitational field of any plausible celestial body \cite{thorne}. To overcome this limitation, a perturbative approach based on the Lewis metric was employed to find a solution with an explicit quadrupole moment, using the Kerr spacetime as seed metric \cite{Frutos1}. The mass quadrupole moment quantifies deviations from the perfect spherical symmetry induced by rotation and matter distribution \cite{quevedo,Gravitation} and becomes especially significant for rapidly rotating neutron stars \cite{oliva} and other compact objects, which independent multipole moments could encode information about their equation of state (EoS), rotation, and internal composition. The quadrupole moment modifies the motion of test particles, affects orbital and epicyclic frequencies, changes the location of orbits and influences both electromagnetic and gravitational wave (GW) signals emitted near compact objects. Consequently, studying quadrupolar corrections to the Kerr geometry is essential for developing realistic models of neutron stars and for probing the multipolar structure of strong gravitational fields. The resulting Kerr-like metric retains a simple analytic form, which makes it well suited for analytical studies such as the determination of the innermost stable circular orbit (ISCO). While exact solutions provide a more complete description of highly deformed sources, approximate metrics often offer a more practical framework for analytical calculations while retaining the essential physical effects associated with rotation and quadrupolar deformations \cite{Frutos2013,Frutos2015}.

In classical mechanics, a particle moving under the influence of gravity can stably orbit a central object at any radius as long as it has sufficient angular momentum that allows it. There is no smallest radius where stable circular orbits cease to exist and the effective potential with its minimum gives us the possibility to determine where this orbit happens. In general relativity the situation changes. The effective potential for curved spacetimes typically presents two extrema, a maximum corresponding to a unstable circular orbit and a minimum to a stable one. As the particle's angular momentum decreases, the two extrema approach each other and finally merge into a single point: the ISCO \cite{moore,bardeen,tsupko,oleg}. Below this point, the effective potential has no extrema, which means that there are not stable or unstable orbits and particles will inevitably fall into the compact object. The ISCO represents the smallest radius at which a test particle can orbit, often interpreted as the inner boundary of the accretion disk surrounding compact objects. It plays an important role in phenomena such as quasi periodic oscillations (QPOs), X-ray emission, and GW generation \cite{novikov,shakura,morsink,abramowicz}. In accreting systems, the ISCO defines the characteristic spatial scale for the innermost region of the accretion flow. Within the standard thin-disk framework, the ISCO dictates the inner edge of the disk provided the stellar surface does not extend beyond this boundary, thereby directly influencing the radiative efficiency and high-energy emission produced in the strong-field regime. For neutron stars, the situation is particularly rich, as the inner disk boundary may be determined by the ISCO, the stellar radius, or magnetospheric interactions \cite{pappas,cipolletta,luk}. Consequently, observational constraints on the inner disk radius can provide valuable bounds on the mass, angular momentum, quadrupole moment, stellar radius, and dense-matter EoS of the compact object. Furthermore, the ISCO is of paramount importance in extreme mass-ratio inspirals (EMRIs), where the transition from stable inspiral to plunge occurs deep within the strong-field region and exhibits strong sensitivity to the multipolar structure of the central object \cite{ryan}.

Naturally, the rotation of the central body affects the particle’s motion and leads to rotating and non-rotating cases \cite{Shibata,suzuki,oleg,tsupko}. The rotation of a central massive body influences the motion of particles orbiting it, which explains why the trajectories of test particles differ in the Schwarzschild and Kerr backgrounds. When the test particle itself possesses spin, it will also influence the particle’s orbit \cite{mathisson,Papapetrou,coripapa,dixon} ---something evident even in the non-rotating Schwarzschild background. Following this idea, even for spinless test particles, the ISCO radius depends on the orientation of the orbital motion in a rotating spacetime such as Kerr. Unlike the Schwarzschild spacetime, which is spherically symmetric, the Kerr geometry is stationary and axisymmetric, with its axial symmetry arising from the rotation of the central object. This rotation gives rise to the phenomenon of \textit{frame dragging}, whereby the angular momentum of the central object induces a dragging of inertial frames in its vicinity. As a consequence, the dynamics of particles near a rotating compact object depend on the orientation of their orbital angular momentum relative to the spin of the central object \cite{bardeen}. For equatorial circular orbits, the two possible configurations are conventionally referred to as co-rotating and counter-rotating orbits, corresponding to orbital angular momentum parallel and antiparallel to the angular momentum of the central object, respectively. These features are also present in approximate Kerr-like spacetimes with a quadrupole moment, where the effects of the quadrupole deformation can be investigated for both non-spinning and spinning test particles \cite{semerak,Frutos2,Solano,suzuki}.

This paper is organized as follows. Section \ref{sec:metric} provides a brief review of the metric used and discusses the relationship among metrics and relativistic multipoles, after naming the limiting cases. Section \ref{sec:spin} focuses on the mathematical treatment to understand the behavior of spinless and spinning test particles in a relativistic spacetime. We explore the relative orientation of a particle’s motion and the angular momentum of the central body and how the ISCO parameters are affected by the dynamics. For spinning particles, we include the choice of a supplementary condition and the derivation of constants of motion, which enable us to classify the particle's behavior and quantify the differences considering the particle's spin. Following Chandrasekhar's approach \cite{Chandrasekhar} to define the effective potential, we then develop a novel method in Section \ref{sec:miem} to construct an ISCO function. This framework allows us to simultaneously determine the ISCO radii, conserved energy $E$, and angular momentum $L_z$ across the various configurations. Section \ref{sec:lit} is devoted to obtaining the limiting cases offered by the ISCO polynomial for solutions encompassed within the metric employed in this work, specifically for the case of spinning particles. Section \ref{sec:iscorel} is dedicated to mathematical relations between the energy and the angular momentum in the ISCO. Finally, Section \ref{sec:conclu} is reserved for conclusions. Additional mathematical details and results from the most general metric are provided in the Appendices section.

Throughout this work we employ geometrical units in which $c=G=1$, and we adopt the metric signature $(-++++)$. We performed the tensor and algebraic calculations using the \textsc{Reduce} computer algebra system. These programs are available upon request.

\section{Metric}\label{sec:metric}

In general relativity, a stationary spacetime is the one where the metric tensor remains constant over time, ensuring that the spacetime's geometric structure does not change as time progresses \cite{carmeli}. The Lewis metric is a general stationary spacetime given by
\begin{equation}
    ds^2 = - V dt^2 + 2 W dt\,d\phi + X dr^2 + Y d\theta^2 + Z d\phi^2 \label{lewis}
\end{equation}
with all metric functions depending on $r$ and $\theta$ \cite{lewis,Frutos1}. The five functions of the metric are then defined as follows
\begin{align}
    V &= V_K \, {\rm e}^{-2 \psi} = - \frac{{\rm e}^{-2 \psi}}{\rho^2} \left( a^2 \sin^2 \theta - \Delta \right)\,, \nonumber \\
    W &= W_K + W_q = \frac{a \sin^2\theta}{\rho^2}\Big[\Delta - \left(r^2 + a^2\right)\Big] + W_q\,, \nonumber \\
    X &= X_K \, {\rm e}^{2\chi} = \frac{\rho^2}{\Delta}{\rm e}^{2\chi}\,, \\ 
    Y &= Y_K \, {\rm e}^{2\chi} = {\rho^2}{{\rm e}^{2\chi}}\,, \nonumber \\
    Z &= Z_K \, {\rm e}^{2\psi} = \frac{{\rm e}^{2\psi}}{\rho^2} \sin^2 \theta\, \Big[ \left(r^2 + a^2\right)^2 - a^2 \Delta \sin^2 \theta \Big] \,, \nonumber
\end{align}
where $V_K, W_K, X_K, Y_K $ and $ Z_K $ are the Kerr metric potentials \cite{kerr}. The functions are $\rho^2=r^2+a^2\cos^2\theta$ and $\Delta=r^2-2Mr+a^2$, as in the Kerr metric. The \textit{Kerr parameter} $a$ is defined as $a\equiv J/M$, where $J$ represents the total angular momentum of the object and $M$ is its mass \cite{Gravitation}. Each function also has additional expressions that need to be clarified. We have the following expansions accompanying the exponential terms
\begin{equation}
\begin{aligned}
    \psi &= q\frac{P_2}{r^3} + 3 M q \frac{P_2 }{r^4}\,, \\ 
    \chi &= q \frac{P_2}{r^3} + \frac{M q}{3 r^4} \left(5 P_2^2 + 5 P_2 - 1\right) \\
    &\quad+ \frac{q^2}{9 r^6} \left(25 P_2^3 - 21 P_2^2 - 6 P_2 + 2\right) 
\end{aligned}
\end{equation}
with $P_2=(3\cos^2\theta-1)/2\,$ and $q$ depicting the Legendre polynomial and the mass quadrupole parameter respectively ---the latter incorporates $q$ up to second order. The additional term $W_q$ is given by
\begin{equation}
    W_q = \frac{J q}{r^4}P^1_{3} \sin{\theta}\,,
\end{equation}
where $ P^1_3 = (5 P_2 + 1) \sin{\theta}$ is an associated Legendre polynomial. The metric function $ W $ ---after considering the perturbative approximation $a^2{\rm e}^{-2\eta}\sim a^2, \, (\mathcal{O}(a^2q)\sim 0) $--- becomes 
\begin{equation}
    W \simeq - 2 \frac{J r}{\rho^2} \, {\rm e}^{- \eta} \, \sin^2{\theta} \, ,
\end{equation}
where
\begin{equation}
    \eta = \frac{q}{2 r^3} (5 P_2 + 1) \, .
\end{equation}

The Kerr-like metric with quadrupole (KLMQ), employed in this article, describes the spacetime of massive, rotating and deformed objects, like rotating neutron stars and black holes. This metric was derived by incorporating the quadrupole moment as a perturbation, valid up to second order in $q$, providing an approximate solution of Einstein's field equations. As special cases, this spacetime includes the Kerr metric \cite{kerr} ($q = 0$), the Erez-Rosen metric \cite{ErezRosen} expanded in Taylor series up to the second order in $q$ (non-rotating mass with a quadrupole moment, $a = 0$), the Lense-Thirring metric \cite{lensethirring,mashhoon1984} (for slow rotation, $a^2 \sim 0$) and the Schwarzschild metric \cite{schwarzschild} ($q = a = 0$) \cite{Frutos2}.

The exact Kerr metric uniquely describes a stationary, axisymmetric vacuum black hole whose multipole structure is completely specified by only two global parameters: the total mass $M$ and the spin parameter $a \equiv J/M$ \cite{carter1971,robinson1975}. In the Geroch--Hansen multipolar expansion \cite{geroch,hansen}, the mass multipoles $M_\ell$ and angular momentum multipoles $S_\ell$ satisfy the concise relation
\begin{equation}
M_\ell + i S_\ell = M (i a)^\ell ,
\end{equation}
which enforces $M_0 = M$, $S_1 = J = M a$, $M_2 = -M a^2$, and sets all independent multipolar degrees of freedom to zero.

To account for structural deformations in realistic compact objects, such as neutron stars or perturbed black holes, approximate Kerr-like spacetimes featuring an independent mass quadrupole moment have been widely investigated \cite{Frutos2013,Frutos2015,ErezRosen,hartle1967,HartleThorne,QuevedoMashhoon1991,Quevedo2011,MankoNovikov1992}. Unlike the exact Kerr solution, where all multipoles are uniquely determined by $M$ and $a$, these deformed metrics allow the quadrupole parameter to vary independently, offering a powerful framework to describe the exterior geometry of non-Kerr compact objects. In particular, the metric employed in this work follows the perturbative approach, incorporating mass quadrupole corrections up to second order. The KLMQ multipoles are then 
\begin{equation}
\begin{aligned}
    M_0=M\,,\,S_1&=Ma\,,\,M_2= -M a^2 + q\\
    P_\ell &= M(ia)^{\ell}\quad \ell>2,
\end{aligned}
\end{equation}
where $P_\ell$ denotes the higher-order multipoles, which coincide with those of the Kerr spacetime to the order of approximation considered. In this deformed Kerr-like spacetime, the mass quadrupole moment $M_2$ is no longer constrained by the Kerr relation $M_2 = -M a^2$, but incorporates an independent quadrupole parameter $q$. 
\section{Motion of a test particle}\label{sec:spin}

\subsection{Spinless particles}

In \cite{Frutos2}, the authors focused on deriving the ISCO equation for the Kerr-like metric that incorporates the mass quadrupole moment $q$. The metric used in the referenced article is identical to the one employed here, up to a minor modification ($W_q=0$). By means of the Chandrasekhar’s method \cite{Chandrasekhar} using Lagrangian and Hamiltonian dynamics for a non-spinning test particle, they derived expressions for the energy and the angular momentum after applying stability conditions for the effective potential. 

\begin{figure}
    \centering
        \includegraphics[width=\linewidth]{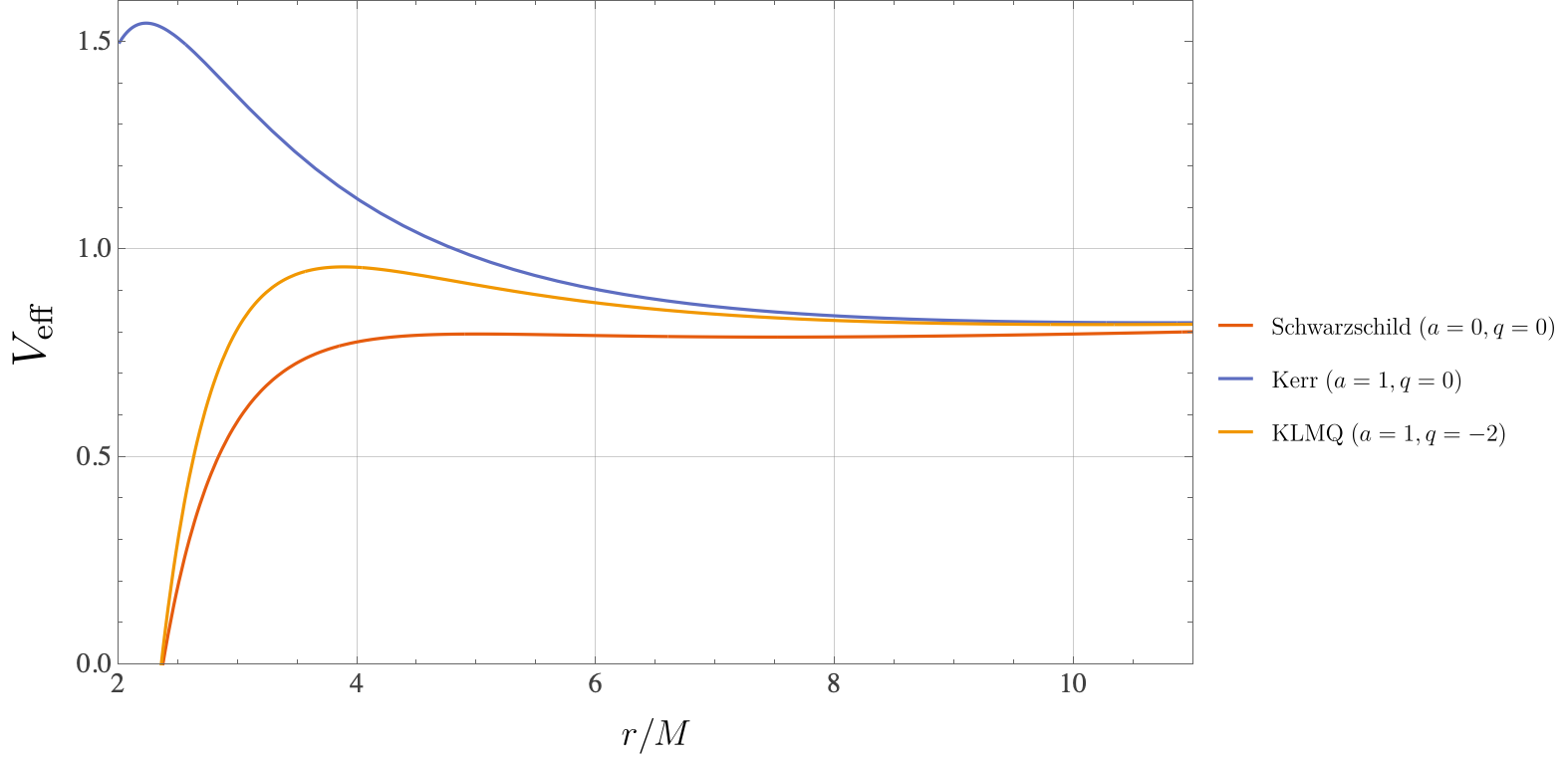}
        \caption{Effective potential $V_{\rm eff}$ for $E=1$ and $L_z=5$ in the Schwarzschild ($a=q=0$), extreme Kerr ($a=M$, $q=0$), and KLMQ ($a=M$, $q=-2$) spacetimes. Units with $M=1$ are used.}
        \label{fig:veff}
\end{figure}

\begin{figure}
    \centering
        \includegraphics[width=\linewidth]{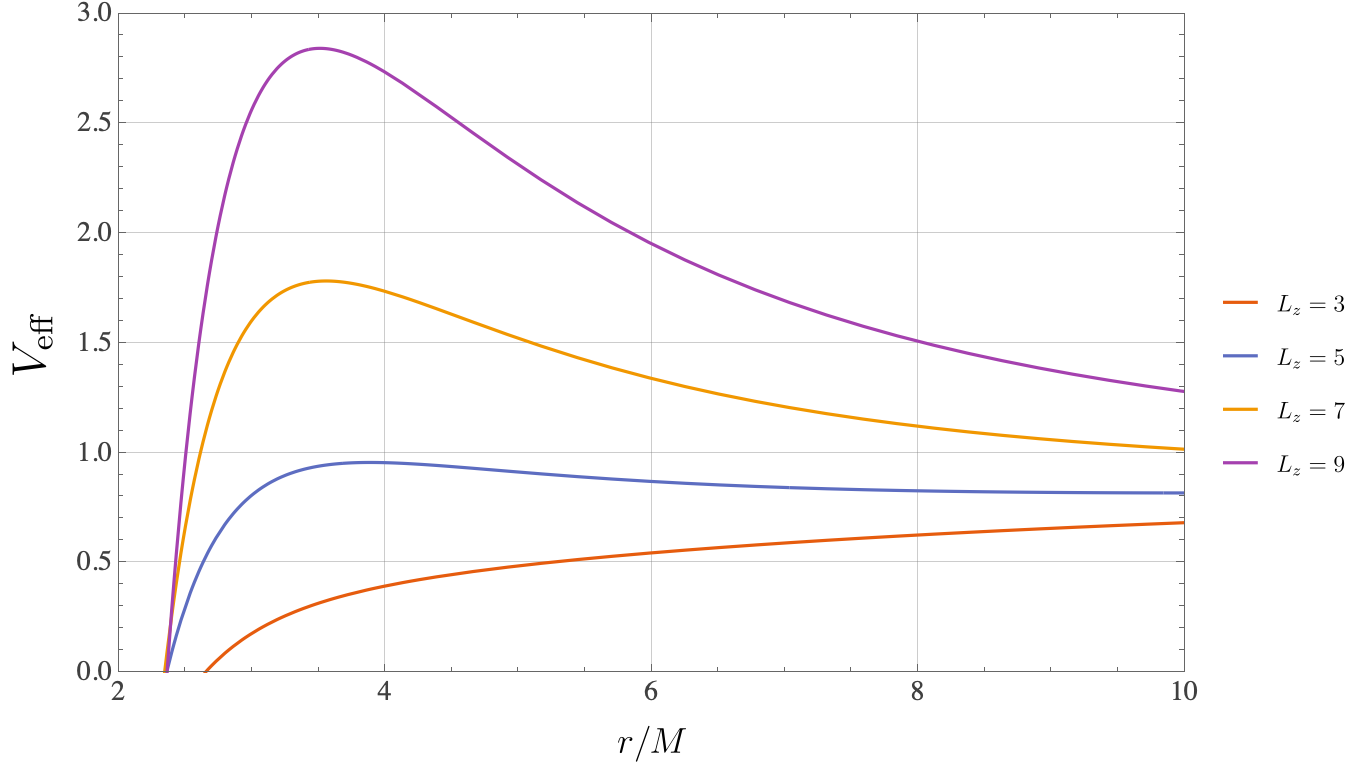}
        \caption{Effective potential $V_{\rm eff}$ of the KLMQ spacetime with $a=M$ and $q=-2$, for $E=1$ and different values of $L_z$, as a function of $r/M$ (units with $M=1$).}
        \label{fig:lz}
\end{figure}

\begin{figure}
    \centering
        \includegraphics[width=\linewidth]{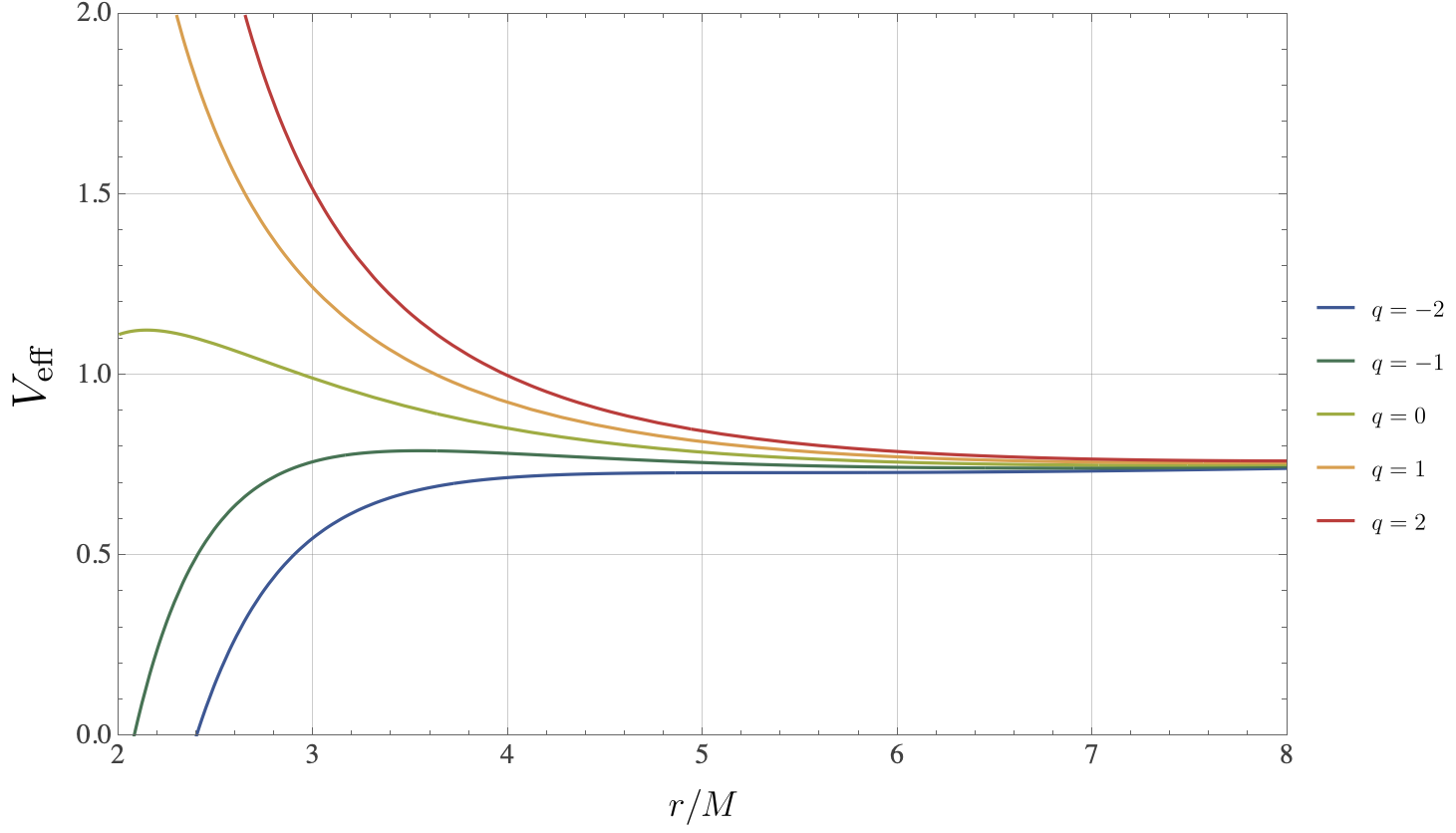}
        \caption{Effective potential $V_{\rm eff}$ for $E=1$, $L_z=4.23$, and $a=M$, for different values of the quadrupole parameter $q$ (units with $M=1$). The sign of $q$ determines whether the central object is oblate ($q<0$) or prolate ($q>0$), directly affecting the profile of the effective potential and the formation of its extrema.}
        \label{fig:klmq}
\end{figure}

The variable change $u=1/r$, widely employed in classical  mechanics, is used to derive the expressions. With the metric components set for particles in the equatorial plane ($\theta=\pi/2$) and the Hamiltonian formalism, the effective potential obtained is
\begin{equation}
\begin{aligned}
    V_{\text{eff}} = -2\epsilon\mathcal{A} + L_z^2\mathcal{B}- 2Mu^3 (L_z - aE)^2 - E^2\mathcal{C} \label{pot}
\end{aligned}
\end{equation}
with
\begin{align*}
    \mathcal{A} &= 1 - 2Mu + a^2u^2 + qu^3 - \frac{1}{2} Mqu^4 + \frac{5}{4} q^2 u^6 \,,\\
     \mathcal{B} &=u^2 + 2qu^5 + \frac{1}{2} Mqu^6 + \frac{11}{4} q^2 u^8 \,, \\
    \mathcal{C} &= 1 + a^2u^2 - \frac{3}{2} Mqu^4 + \frac{3}{4} q^2 u^6 \,, 
\end{align*}
where $\epsilon$ represents the parameter that determines the nature of the geodesic in the spacetime; $\epsilon=-1$ for time-like geodesics, $\epsilon=0$ for light-like geodesics and $\epsilon=1$ for space-like geodesics. The behavior of the effective potential is illustrated in Figs. \ref{fig:veff}-\ref{fig:klmq}. Importantly, from this potential we can also recover limiting cases such as the Schwarzschild and Kerr metrics. For simplicity, we choose an extreme black hole with $a=M$ and we set $M=1$, $E=1$, and $L_z=5$. Finally, we set $\epsilon=-1$, corresponding to massive particles \cite{carroll}. As shown in Figure $\ref{fig:veff}$, for $a=1$ and $L_z=5$, extremal Kerr develops its potential barrier very close to the black hole ($r\approx2.2$). The oblate quadrupole correction ($q=-2$) shifts this same barrier outward (to $r\approx3.9$) and reduces its height, making it visible within the plotted range and less pronounced than in pure Kerr. Schwarzschild, by contrast, lies just above its own critical angular momentum threshold, so its extremum structure nearly degenerates into a single inflection point. At infinity, the effective potential for each metric tends to a constant which is independent of the value of the parameters of the test particle and of the gravitational source ---e.g. the rotation of the central body, the quadrupole moment, or even the particle's spin \cite{quevedo2}. Rather than creating a trapping potential well where none previously existed, the quadrupole moment in this regime primarily displaces and flattens the centrifugal barrier associated with the black hole's rotation. Table \ref{tab:veff} presents the local maxima and minima corresponding to the curves plotted in Figure \ref{fig:veff} for the three backgrounds for that specific combination of parameters.

\begin{table}
    \begin{ruledtabular}
    \begin{tabular}{ccc}
         & Maximum (unstable) & Minimum (stable) \\
        \hline
        S        & $r\approx5.00\to V_{\text{eff}}\approx0.80$ & $r\approx7.50\to V_{\text{eff}}\approx0.79$ \\
        K          & $r\approx2.23\to V_{\text{eff}}\approx1.55$ & $r\approx10.77\to V_{\text{eff}}\approx0.83$ \\
        KLMQ  & $r\approx3.88\to V_{\text{eff}}\approx0.96$ & $r\approx10.23\to V_{\text{eff}}\approx0.82$
    \end{tabular}
    \end{ruledtabular}
    \caption{Extrema of the effective potential for the Schwarzschild (S), Kerr (K), and KLMQ ($q=-2$) metrics for spinless particles in an extreme black hole $a=M$.}
    \label{tab:veff}
\end{table}

In general, the effective potential for curved spacetimes presents two extrema. As the particle's angular momentum decreases, these two extrema approach each other and eventually merge at a single point, corresponding to the ISCO. We can thus plot the effective potential for various values of $L_z$ to illustrate this behavior. As illustrated in Figure \ref{fig:lz}, the maximum and minimum (the latter located outside the plotted region) progressively approach each other as the angular momentum is decreased. The presence of an additional parameter ---the mass quadrupole moment--- implies that the specific value of $L_z$ at which the ISCO occurs depends strongly on the chosen value of $q$. Numerically, the values of $L_z$ corresponding to the ISCO could be obtained by varying $q$. Although the choice of $q = -2$ is arbitrary and adopted primarily for visualization purposes, since it makes the quadrupolar effects clearly visible in the plots, the quantitative values obtained for such a large deformation should be regarded as illustrative, given the perturbative character of the metric. We shall maintain this value throughout the remainder of this work. For this specific value of $q$, we find that the angular momentum at which the maximum and minimum merge is $L_z \approx 4.23$. 

We can also plot the effective potential for the extreme Kerr case while varying the quadrupole parameter. Depending on the sign of $q$, different mass distributions are obtained for the rotating central object. Negative values of the quadrupole parameter represent oblate configurations, whereas positive values describe prolate ones \cite{bini,adrian}. 

Figure~\ref{fig:klmq} shows that, for a fixed angular momentum $L_z=4.23\,$, the sign of $q$ does not primarily control the depth of the minimum of $V_{\mathrm{eff}}$, which persists for every value of $q$ considered and remains located far from the compact object ($r\gtrsim10$) with only a mild dependence on $q$ ---it is deeper for oblate ($q<0$) configurations, consistent with the extra attractive contribution expected from an equatorial mass excess. Instead, the sign of $q$ controls the visibility and location of the potential barrier associated with the unstable circular orbit. For oblate configurations ($q<0$),
this local maximum appears well outside the black hole. For $q\geq0$
(spherical and prolate configurations), the corresponding barrier recedes toward the horizon and lies outside the plotted range, exactly as in the pure Kerr case ($q=0$), whose barrier sits at $r\approx2.2$. This does not imply that oblate configurations are uniquely capable of
capturing particles, since a genuine potential well is present for every sign of $q$. Rather, the quadrupole moment primarily reshapes the region near the unstable orbit: an oblate deformation pushes the centrifugal barrier outward and flattens it, exposing a structure that,
for a spherical or prolate source with the same angular momentum, would remain hidden much closer to the black hole.

Proceeding with the effective potential analysis, the expression (\ref{pot}) was then differentiated twice to identify the values of $r$ where the orbit is stable. With the substitution $x=aE-L_z$, the authors rewrote the effective potential and its derivatives, imposed the circular-orbit conditions and, after some algebraic manipulation, obtained the following expression for the energy,
\begin{equation}
\begin{aligned}
      E^2&=\frac{-2\epsilon}{\mathcal{Z}_{\pm}}\bigg[(1-2Mu)\left(1-2Mu\pm2au\sqrt{Mu}\right)\\
      &\quad+\left(aJ-\frac{1}{2}q\right)u^3+\frac{25}{2}Mqu^4+\frac{29}{2}q^2u^6 \bigg] \label{e}
\end{aligned}
\end{equation}
with 
\begin{equation*}
    \mathcal{Z}_{\pm}=\Big(1-3Mu+\dfrac{33}{2}Mqu^4+11q^2u^6\Big) \pm2au\sqrt{Mu}\,.
\end{equation*}
The roots of $\mathcal{Z}_{\pm}$ represent the radii $r_{\gamma \pm}$ of photon orbits, at which the energy in (\ref{e}) diverges. This divergence reflects the fact that only massless particles can follow circular orbits at these radii \cite{Solano}.

The angular momentum in the $z$-direction was also derived and is defined as follows, where $\mathbb{Z}=u^2\mathcal{Z}_{+}\mathcal{Z}_{-}$ 
\begin{equation}
\begin{aligned}
    L^2_z&=\frac{-2\epsilon u}{\mathbb{Z}}\bigg[M - 3M^2u + \left( 2aJ - \frac{3}{2} q \right) u^2 \\
    &\quad+ \left( 6J^2 - \frac{5}{2}Mq \right) u^3 + aJ \left( a^2 - 12M^2\right) u^4 \\
    &\quad + \left( 5a^2J^2 + 6q^2 \right) u^5 \pm 2Ju \sqrt{Mu}\times \\
    &\quad\Big( a^4 u^4 - 2aJ u^3 + 4a^2 u^2 - 6Mu + 3\Big)\bigg]\,.
\end{aligned}
\end{equation}

After changing again $u=1/r$, the final ISCO equation is obtained as a function of $r$
\begin{equation}
\begin{aligned}
    \mathcal{P}&=Mr^5-9M^2r^4+3\left(6M^3-aJ+\frac{1}{2}q\right)r^3\\
    &\quad-\bigg(7J^2-\frac{29}{2}Mq\bigg)r^2-\frac{33}{2}q^2\pm6Jr\sqrt{Mr}\Delta \label{isco}
\end{aligned}
\end{equation}
which generalizes the Schwarzschild and Kerr ISCO expressions. By setting $\mathcal{P}=0$ to determine its roots, we obtain as solutions the radii at which the particle would orbit in the ISCO.

The derivation is a non-trivial extension of the well-known Kerr case, including quadrupole moment terms up to second order. Equation ($\ref{isco}$) explicitly includes the quadrupole parameter. Given $M$, $a$ and $r_{\text{ISCO}}$, one can, in principle, solve for $q$, allowing an indirect measurement of a compact object’s quadrupole moment.

Although $\mathcal{P}$ admits several roots, not all of them are physically meaningful; only those lying outside the horizon and satisfying the stability conditions correspond to an ISCO. As in the case of the Kerr metric, we obtain two solutions for the ISCO radius, a consequence of the $\pm$ sign in the corresponding equation. At this point, a crucial factor enters the analysis: the relative orientation of the particle’s orbital motion with respect to the rotation of the compact object.

\begin{figure}
    \centering
    \includegraphics[width=\linewidth]{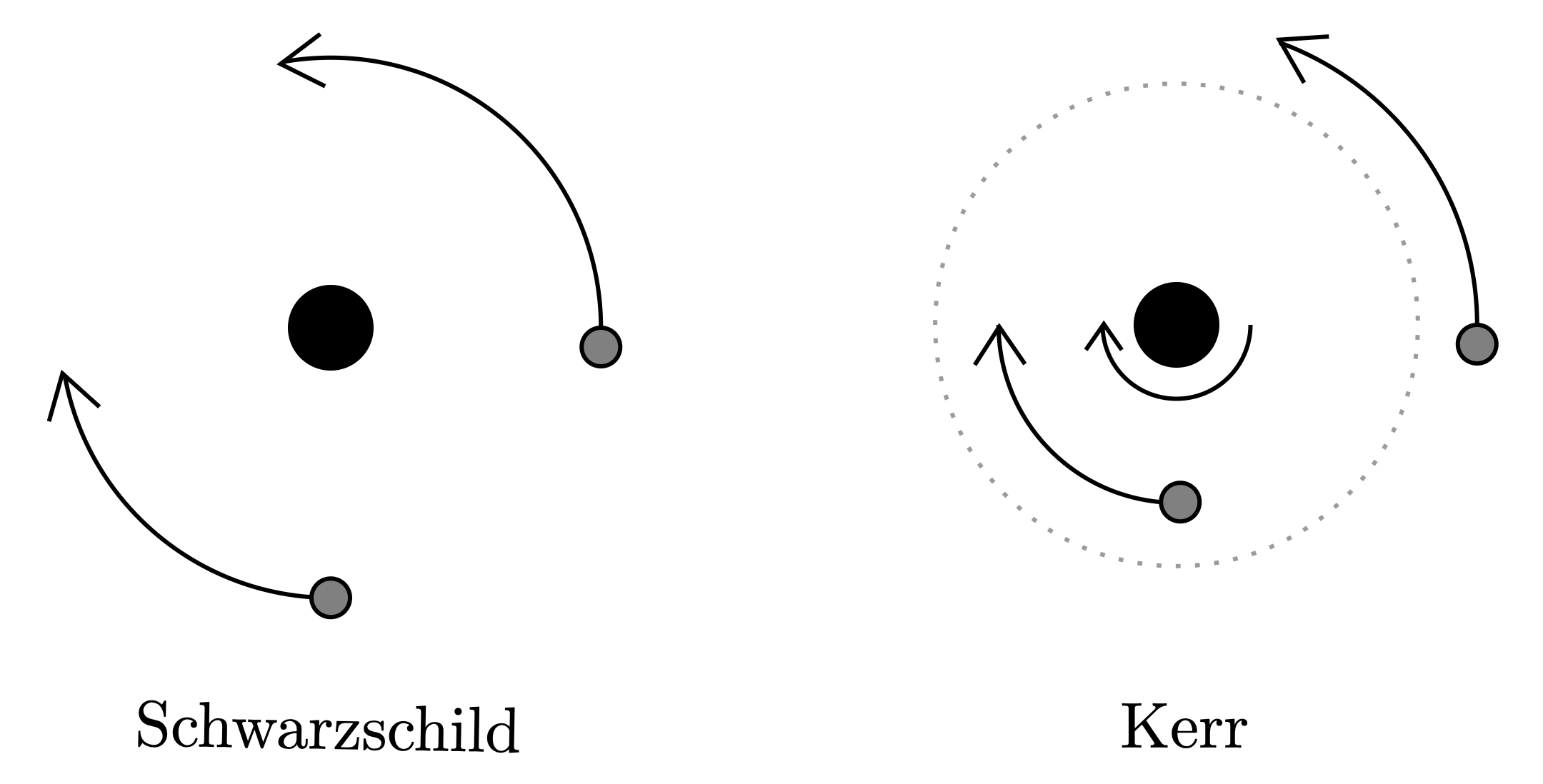}
    \caption{A not-to-scale representation of how the angular momentum of the central body affects the orbit of a particle. Co-rotation and counter-rotation do not influence a particle in the Schwarzschild background, but in the Kerr spacetime, one can notice that the orbit shifts from the $6M$ limit (dotted line) inward or outward.}
    \label{fig:rotation}
\end{figure}

The ISCO for spinless test particles in Schwarzschild and Kerr spacetimes exhibits different behaviors depending on the relative orientation of the particle’s motion and the angular momentum of the central body. The $r_{\text{ISCO}}$ for Schwarzschild is $6M$ independently of the orbital direction. Co-rotating and counter-rotating cases correspond to parallel and anti-parallel alignments, respectively, between the particle's and the compact object's angular momentum. The rotation of the central mass ---e.g., a extreme Kerr ($a=M$) black hole--- significantly influences the ISCO radius as illustrated in Figure $\ref{fig:rotation}$: in the case of co-rotation, the ISCO shifts inward to radii smaller than $6M$, while in counter-rotation it moves outward beyond $6M$. Co-rotating scenario gives us  $r_{\text{ISCO}}=M$ and for counter-rotating orbit we have  $r_{\text{ISCO}}=9M$ \cite{bardeen,oleg,tsupko}.

This behavior applies to an approximate Kerr-like spacetime with mass quadrupole. The necessary condition for recovering the seed metric is $q=0$, which allows us to verify that the ISCO equation yields, as its roots, the well-known Kerr results. When we extend the analysis and take into account the massive quadrupole moment, we find that this parameter also affects the radial distance at which a test particle can orbit. Since the equation does not admit closed-form solutions in general, it must be solved numerically for given values of $a$ and $q$ in order to determine with greater precision the deviation in $r_\text{ISCO}$ induced by the quadrupole.

\subsection{Spinning particles}

The dynamics of spinning test bodies in general relativity are governed by the Mathisson-Papapetrou-Dixon (MPD) formalism, originally conceived by Mathisson \cite{mathisson} to describe extended objects in curved spacetimes, further developed by Papapetrou \cite{Papapetrou,coripapa}, and later refined into a rigorous multipolar framework by Dixon \cite{dixon}. The resulting equations of motion of a spinning test particle in a relativistic curved spacetime are given by 
\begin{equation}
\begin{aligned}
    \frac{D}{D\tau}p^{\mu}&=-\frac{1}{2}R^{\mu}_{\nu\sigma\eta}v^{\nu}S^{\sigma\eta}\,,\\
     \frac{D}{D\tau}S^{\mu\nu}&=  p^{\mu}v^{\nu}-p^{\nu}v^{\mu} \label{papa}
\end{aligned}
\end{equation}
where $\tau,p^{\mu},R^{\mu}_{\nu\sigma\eta},v^{\nu},S^{\mu\nu}$ represent the affine parameter of the orbit, the 4-momentum, the Riemann tensor, the 4-velocity of the particle and the spin tensor, respectively. The notation $D/D\tau$ stands for the covariant derivative along the trajectory of the particle. The term on the right of the first equation is called \textit{spin-curvature interaction} \cite{carmeli}. Additionally, for the sake of completeness, it is known that the 4-velocity is defined as $v^{\mu}=dx^{\mu}/d\tau\,.$

To fully determine the equations of motion, it is necessary to apply additional conditions that define the line corresponding to the path of the test particle. The supplementary condition
\begin{equation}
   p_\mu S^{\mu\nu}=0\,
\end{equation}
prevents obtaining a nonphysical helical motion even in a flat spacetime \cite{tulczyjew,carmeli}. 

Parameters of ISCO like radius, total angular momentum, energy and orbital angular
frequency have been widely investigated in Schwarzschild and Kerr spacetimes for spinning test particles \cite{Shibata,oleg,suzuki}. If the metric admits Killing vectors $\xi^{\mu}$ that satisfy the Killing equation $\nabla_{(\mu}\xi_{\nu)}=\nabla _\mu\xi_{\nu}+\nabla _\nu\xi_{\mu}=0$, then scalar conserved quantities associated with each symmetry can be derived using the relation
\begin{equation}
    Q=\xi^{\mu}p_{\mu}-\frac{1}{2}\xi_{\mu;\nu}S^{\mu\nu} \label{quan}
\end{equation}
where $\xi_{\mu;\nu}=\nabla _\nu\xi_{\mu}=\partial_{\nu}\xi_{\mu}-\Gamma^{\sigma}_{\nu\mu}\xi_{\sigma}$ represents the covariant derivative of the Killing vector. We can express the two covariant Killing vectors of the metric ($\ref{lewis}$) using the definition of the functions in terms $(t, r, \theta, \phi)$ coordinates.
The time-like Killing vector is 
\begin{equation}
\begin{aligned}
    \xi^{(t)}_\mu &= \frac{{\rm e}^{-2 \psi}}{\rho^2} \left( a^2 \sin^2 \theta - \Delta \right) \delta^t_\mu \\
      &\quad+ \left(\frac{a \sin^2 \theta}{\rho^2} \Big[ \Delta - \left(r^2 + a^2\right) \Big] +W_q\right)\delta^\phi_\mu \label{k1}
\end{aligned}
\end{equation}
and the space-like axial Killing vector is given by
\begin{equation}
\begin{aligned}
    \xi^{(\phi)}_\mu &=  \left(\frac{a \sin^2 \theta}{\rho^2} \Big[ \Delta - \left(r^2 + a^2\right) \Big]+ W_q \right) \delta^t_\mu \\
      &\quad+ \frac{{\rm e}^{2\psi}}{\rho^2} \sin^2 \theta\, \Big[ \left(r^2 + a^2\right)^2 - a^2 \Delta \sin^2 \theta \Big] \delta^\phi_\mu\,. \label{k2}
\end{aligned}
\end{equation}
It is straightforward to show that these Killing vectors are slight modifications of those derived from the Kerr metric. We can then express them as
\begin{align*}
    \xi^{(t)}_\mu &= V_K\,{\rm e}^{-2 \psi}\delta^t_\mu + (W_K+W_q)\delta^\phi_\mu\,,\\
    \xi^{(\phi)}_\mu &= (W_K+W_q)\delta^t_\mu + Z_K\,{\rm e}^{2\psi}\delta^\phi_\mu
\end{align*}
where the metric components represent the unperturbed components of the Kerr metric.

Having defined the Killing vectors, we can now apply relation (\ref{quan}) to compute the energy and the $z$- component of total angular momentum, which correspond to the symmetries associated with each vector, respectively. By employing it and using the Killing vectors defined in (\ref{k1}) and (\ref{k2}), we determine that the energy of a spinning particle in the Kerr-like spacetime with mass quadrupole and the particle's total angular momentum in the $z$- direction ---without any loss of generality--- are respectively 

\begin{equation}
\begin{aligned}
    -E &= {\rm e}^{-2\psi}\,\frac{\mathcal{K}_1}{2\rho^{6}}\,S^{tr}
    -{\rm e}^{-2\psi}\,\frac{\mathcal{K}_2\sin2\theta}{2\rho^{6}}\,S^{t\theta}
    \\
    &\quad+ \left[\frac{J\sin^{2}\theta\,(r^2-a^2\cos^2\theta)}{\rho^{4}}
    +\frac{24Jq(P_4-P_2)}{7r^{5}}\right]S^{r\phi}
    \\&\quad+\left[-\frac{Jr\,(a^{2}+r^{2})\sin2\theta}{\rho^{4}}
    +\frac{6Jq\,P_3\sin\theta}{r^{4}}\right]S^{\theta\phi}+p_t\,,
\end{aligned}
\end{equation}
\begin{equation}
\begin{aligned}
    L_z &= {\rm e}^{2\psi}\,\frac{\left(r\,\mathcal{K}_3-2a^{2}\sin^{2}\theta\,\mathcal{K}_4\right)\sin^{2}\theta}{2\rho^{6}}\,S^{r\phi}
    \\
    &\quad+{\rm e}^{2\psi}\,\frac{\left((a^2+r^2)^4+\mathcal{K}_5\sin^2\theta\right)\sin2\theta}{2\rho^{6}}\,S^{\theta\phi}
    \\&\quad-\left[\frac{J\sin^{2}\theta\,(r^2-a^2\cos^2\theta)}{\rho^{4}}
    +\frac{24Jq\,(P_4-P_2)}{7r^{5}}\right]S^{tr}
    \\
    &\quad+\left[\frac{Jr\,(a^{2}+r^{2})\sin2\theta}{\rho^{4}}
    -\frac{6Jq\,P_3\sin\theta}{r^{4}}\right]S^{t\theta}+p_{\phi}\,,
\end{aligned}
\end{equation}
where
\begin{equation*}
\begin{aligned}
\mathcal{K}_1 &= 3q\,(1-3\cos^{2}\theta)\,r\,(r+2M)+2M\left(r^4-a^4\cos^4\theta\right),\\
\mathcal{K}_2 &= r\left[2a^2M\rho^2+3qr(M+r)\right],\\
\mathcal{K}_3 &= 2\,(a^2+r^2)^3+3q\,(3\cos^{2}\theta-1)\,r^2\,(r+4M),\\
\mathcal{K}_4 &= M\rho^2\,(r^2-a^2\cos^2\theta)\\
    &\quad+r\left[3(a^2+r^2)^2-3a^2(a^2+r^2)\sin^2\theta+a^4\sin^4\theta\right],\\
\mathcal{K}_5 &= -a^2(a^2+r^2)^2\left[3(a^2+r^2)-4Mr\right]+3qr^4(3M+r)\\
&\quad+a^4\,\Delta\,\sin^2\theta\left[3(a^2+r^2)-a^2\sin^2\theta\right]
\end{aligned}
\end{equation*}
and $P_2,P_3$ and $P_4$ as Legendre polynomials.

\begin{figure}
    \centering
    \includegraphics[width=\linewidth]{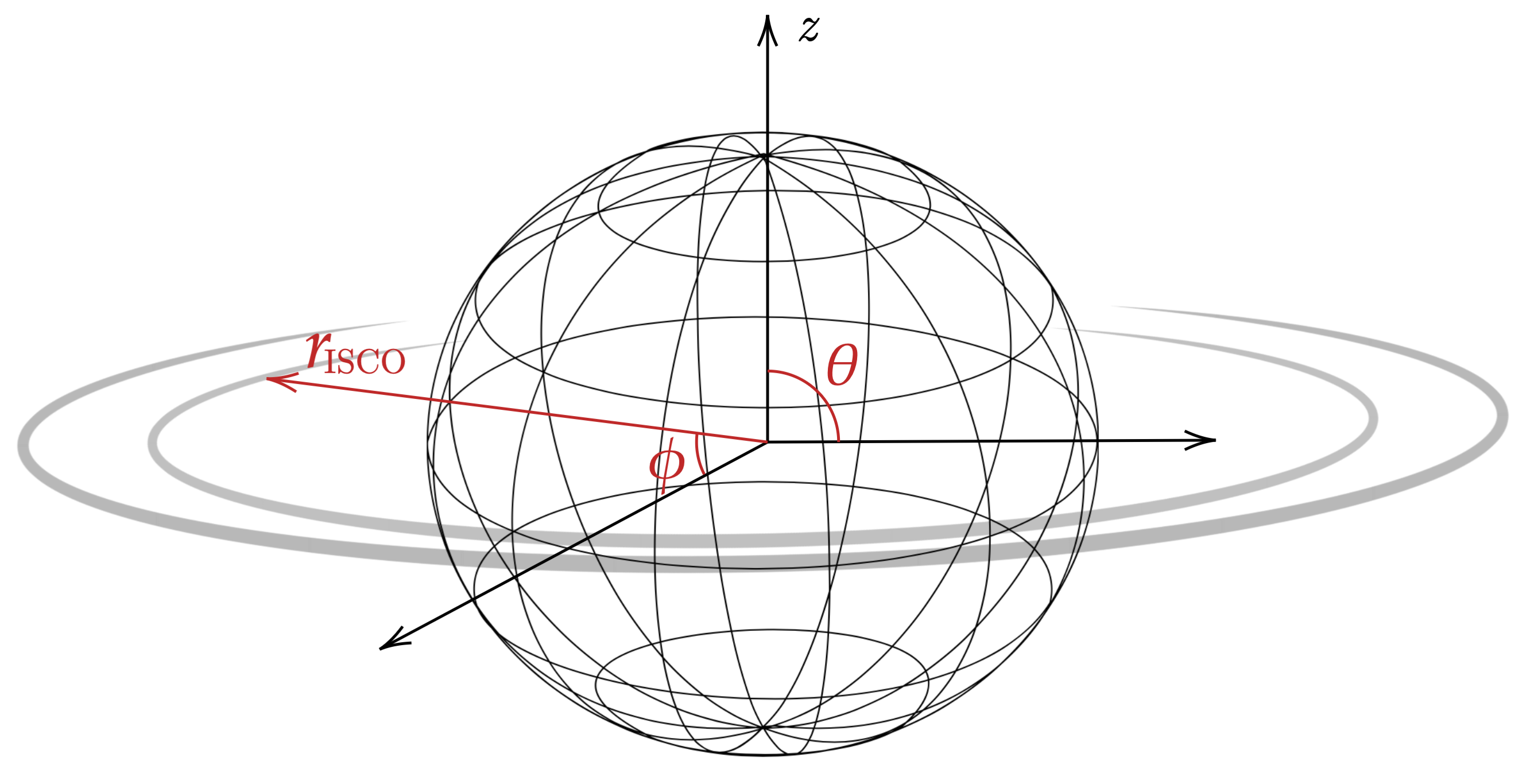}
    \caption{Motion restricted to the equatorial plane ($\theta = \pi/2$). The ISCO represents the smallest radius at which a test particle can orbit; one can interpret it as the inner boundary of accretion disks.}
    \label{fig:equatorial}
\end{figure}

These expressions represent the general conserved quantities for an approximate Kerr-like spacetime with mass quadrupole. These expressions, however, are too involved to be used directly in further calculations. A considerable simplification is achieved by restricting the motion to a particular plane. For simplicity and in analogy with the spinless case, we can restrict the discussion to a particular motion in the equatorial plane ($\theta=\pi/2$) as illustrated in Figure $\ref{fig:equatorial}$, although the motion of a spinning particle is in general not confined to it \cite{bartelmann,carmeli}. 

Conserved quantities are then easier to handle for both Kerr and the Kerr-like metric with quadrupole. Since the metric was obtained by perturbing the Kerr metric as seed metric, it is natural that results of the Kerr spacetime can be recognized in the conserved quantities. By calculating conserved quantities such as energy and total angular momentum for the Kerr-like metric with mass quadrupole, we can identify their counterparts in the unperturbed Kerr background, which is essential for analyzing how the perturbations modify the physical properties of the spacetime and adjust it to a more realistic scenario. 

The expressions differ from the general ones presented before since they are now evaluated at $\theta=\pi/2$. The expression accompanying all metric functions are then written as
\begin{equation}
\begin{aligned}
    \bar{\psi}&=-\frac{q}{2r^4}\left(3M+r\right)\,, \\
    \bar{\chi}&=-\frac{q}{8 r^6} \left( 4 {r^3} + 6 {M}{r^2} + 3 {q} \right)\,,\\
    \bar{\eta}&=-\frac{3q}{4r^3} \,,
\end{aligned}
\end{equation}
where the bar denotes evaluation at $\theta=\pi/2$. We first calculated the conserved quantities in the unperturbed Kerr background to provide a clearer understanding of the differences that arise when using an approximate Kerr-like metric with a mass quadrupole moment, aiming to bring the results closer to those of real compact objects that experience flattening due to rotation. The results for the Kerr solution and their counterparts in the approximate spacetime are shown below.

\begin{equation}
    \begin{aligned}
        -\bar{E}&={\rm e}^{-2\bar{\psi}}\Bigg[\frac{3q\left(2 M +  r\right) }{2 r^{5}} +\frac{M}{r^2}\Bigg]S^{tr}\\
        &\quad+ p_{t}+ \left(\frac{J}{r^{2}}+\frac{3Jq}{r^5}\right)S^{r\phi}, \label{energy}
    \end{aligned}
\end{equation}

\begin{equation}
    \begin{aligned}
        \bar{L}_{z}&={\rm e}^{2\bar{\psi}}\bigg[-\frac{3q(4M+r)}{2r^3}+r-\frac{Ma^2}{r^2}\bigg]S^{r\phi}\\
        &\quad+p_{\phi}-\left(\frac{J}{r^2}+\frac{3Jq}{r^5}\right)S^{tr}\,.\label{momentum}
    \end{aligned}
\end{equation}

It is important to note that these conserved quantities are not strictly perturbations in the sense of simply taking the original Kerr quantities and adding a term. Rather, some terms from the Kerr expressions remain intact, while others are multiplied by the exponential function, along with the addition of entirely new, independent terms. Nonetheless, it is easy to observe that when $q=0$, we recover the Kerr results ---the necessary condition to return the Kerr-like metric with mass quadrupole metric to its seed metric.

Equations (\ref{energy}) and (\ref{momentum}) lead to the following observations. First, we observe that the Schwarzschild energy for a spinning test particle appears explicitly in the overall result. This energy, derived from the static (non-rotating) case, is the component modified by the quadrupolar perturbation, whereas the rotational terms  inherited from the Kerr metric remain unaffected. This behavior is a direct consequence of the chosen approximation: in this regime, the rotational terms generally do not couple with the quadrupolar components. Within this approximation, terms linear in $q$, the couplings $Mq$ and $Jq$, and the term $q^2$ are retained, whereas higher-order couplings involving the quadrupole, such as $M^2q$, $a^2q$, $aq^2$, and $Mq^2$, are neglected. In the angular momentum, by contrast, the quadrupole parameter enters through rotational terms, that couple $q$ to the spin of the central body. Hence, within this approximation, $L_z$ is affected by the quadrupole deformation more directly than $E$. 

We now require a new effective potential that explicitly incorporates the spin magnitude, enabling us to carry out the full mathematical treatment to quantify the small-spin corrections to the ISCO parameters. Physically, the spin of a massive particle is naturally interpreted as a spatial quantity in the particle's instantaneous rest frame, i.e., in the three-dimensional local rest space orthogonal to its four-momentum \cite{semerak,compere,timogiannis}. To describe spin in a curved spacetime while retaining the familiar local Lorentzian structure of special relativity, one introduces a local orthonormal tetrad at each spacetime point. The tetrad formalism provides a locally inertial frame in which tensorial quantities can be expressed in terms of components referred to a locally Minkowskian basis. In particular, the spin tensor can be projected onto this tetrad, allowing its spatial components to be interpreted in the particle's local rest frame. We derive then the following expressions for the co-tetrad associated with the Kerr-like metric 
\begin{equation}
    \begin{aligned}
        e^{(0)} &= \frac{\sqrt{\Delta}}{\rho}\left( {\rm e}^{-\psi} \, dt-a{\rm e}^{\psi-\eta}\sin^2\theta \, d\phi \right)\,,\\
        e^{(1)} &= \frac{\rho \, {\rm e}^{\chi}}{\sqrt{\Delta}} \, dr \,,\\
        e^{(2)} &= \rho \, {\rm e}^{\chi} \, d\theta\,,\\
        e^{(3)} &= \frac{\sin\theta}{\rho}\Big[-a {\rm e}^{-(\psi+\eta)} \, dt+\left(r^2+a^2\right){\rm e}^{\psi} \, d\phi\Big]\,,
    \end{aligned}
\end{equation}
which satisfies
\begin{equation}
\label{apmetric}
 ds^2 = -\left(e^{(0)}\right)^2+\left(e^{(1)}\right)^2+\left(e^{(2)}\right)^2+\left(e^{(3)}\right)^2 .
\end{equation}
Substituting the tetrads into (\ref{apmetric}), we obtain the line element (\ref{lewis}) due to $ a^2{\rm e}^{-2\eta} \sim a^2 $.

To work with vectors (such as the four-velocity $u^\mu$), we require the dual co-frame version. The corresponding vector tetrad $e_{(a)}=e_{(a)}^\mu\partial_\mu,$ is obtained from $e^{(a)}_\mu e_{(b)}^\mu=\delta^{(a)}_{(b)}.$ The non-vanishing components are
\begin{equation}
 \begin{aligned}
 e_{(0)}^t&=\frac{\rho\,(r^2+a^2)\,{\rm e}^\psi}{\Lambda\sqrt{\Delta}}, 
 \\
 e_{(0)}^\phi&=\frac{a\,\rho\,{\rm e}^{-(\psi+\eta)}}{\Lambda\sqrt{\Delta}},
 \\
 e_{(1)}^r&= \frac{\sqrt{\Delta}}{\rho}{\rm e}^{-\chi},
  \\
 e_{(2)}^\theta&=\frac{{\rm e}^{-\chi}}{\rho},
 \\
 e_{(3)}^t &=\frac{a\,\rho\,{\rm e}^{\psi-\eta}\sin\theta}{\Lambda},
  \\
 e_{(3)}^\phi&=\frac{\rho\,{\rm e}^{-\psi}}{\Lambda\sin\theta}.
\end{aligned}    
\end{equation}   
where $\Lambda=(r^2+a^2)-a^2\sin^2\theta\,e^{-2\eta}$.

The antisymmetric tensor $S^{\mu\nu}$ that satisfies the Tulczyjew-Dixon spin supplementary condition $p_\mu S^{\mu\nu} = 0$ \cite{carmeli,tulczyjew} actually possesses only three physical degrees of freedom (since spin is a 3D vector in the rest frame). The standard approach to capturing this feature is to construct the dual vector via the Levi-Civita symbol \cite{Shibata}. The spin four-vector is defined by
\begin{equation}
 s^{(a)}=-\frac{1}{2m}\epsilon^{(a)}_{\phantom{(a)}(b)(c)(d)}u^{(b)}S^{(c)(d)}
\end{equation}
where $u^{(a)}=p^{(a)}/m$. We already know that the quantity $p_\mu p^\mu=-m^2$ remains conserved along the orbit, and we denote it as the particle's mass squared \cite{Shibata}. The test-particle approximation requires $m \ll M$, which we assume throughout. The inverse relation is
\begin{equation}
 S^{(a)(b)}=m\epsilon^{(a)(b)}_{\phantom{(a)(b)}(c)(d)}u^{(c)}s^{(d)}.
\end{equation}
This construction automatically satisfies the Tulczyjew--Dixon supplementary condition $p_{(a)}S^{(a)(b)}=0,$ or equivalently $ p_\mu S^{\mu\nu}=0.$ We first consider an arbitrary value of the polar coordinate $\theta$, but retain the spin orientation used in the equatorial construction. Thus the spin is chosen along the $(2)$ tetrad direction
\begin{equation}
 s^{(a)}=\left(0,0,-\frac{s}{m},0\right),
\end{equation}
where $s$ denotes the total spin angular momentum of the particle and is fixed by the spin-magnitude condition $ 2s^2 = S^{\mu\nu}S_{\mu\nu}\,.$ Consequently $s$ carries the same dimensions as the black hole's spin angular momentum (i.e. $[s]=M^2$ in geometrized units), rather than the dimensions of $M$ often assigned to the specific spin used elsewhere in the literature \cite{oleg,tsupko}. With this choice,
\begin{equation}
\begin{aligned}
 S^{(0)(1)}&=-s\,u^{(3)},\\
 S^{(0)(3)}&=s\,u^{(1)},\\
 S^{(1)(3)}&=s\,u^{(0)},
\end{aligned} 
\end{equation}
whereas $S^{(0)(2)}=S^{(1)(2)}=S^{(2)(3)}=0.$ The coordinate and tetrad components of the spin tensor are related by $ S^{\mu\nu}=e_{(a)}^\mu e_{(b)}^\nu S^{(a)(b)}.$ Thus the nonvanishing coordinate components of the spin tensor are therefore
\begin{equation}
\begin{aligned}
 S^{tr}&=-\frac{s {\rm e}^{-\chi}}{m\rho\sin\theta}\,\frac{p_\phi}{1+\varphi},\\
 S^{t\phi}&=\frac{s {\rm e}^{-\chi}}{m\rho\sin\theta}\,\frac{p_r}{1+\varphi}, \\
 S^{r\phi}&=-\frac{s {\rm e}^{-\chi}}{m\rho\sin\theta}\,\frac{p_t}{1+\varphi}.
\end{aligned}   
\end{equation}
with $\varphi=a^2\sin^2\theta\,(1-{\rm e}^{-2\eta})/\rho^2$. Notice that these expressions are valid for arbitrary $\theta$ under the particular assumption that the spin vector has only a $(2)$ component in the orthonormal tetrad. Since we are working in the equatorial plane, we find that the spin direction of a spinning particle is always perpendicular to the equatorial plane \cite{Shibata}. Hence, the three non-vanishing contravariant components of the spin tensor, after computing them in the equatorial plane ($\varphi\approx-\frac{3}{2}a^2qu^5\to 0$), are
\begin{equation}
\begin{aligned}
 S^{tr}&=-\frac{s{\rm e}^{-\bar{\chi}}}{mr}\,p_\phi,\\
 S^{t\phi}&=\frac{s{\rm e}^{-\bar{\chi}}}{mr}\,p_r,\\
 S^{r\phi}&=-\frac{s{\rm e}^{-\bar{\chi}}}{mr}\,p_t.
\end{aligned}   
\end{equation}
These components of the spin tensor demonstrate a direct influence of the mass quadrupole moment up to second order via the function $\chi$.

With this, we are now able to directly incorporate the spin magnitude of the particle into the conserved quantities obtained previously in equations (\ref{energy}) and (\ref{momentum}). The spin magnitude also allows us to define another type of co-rotation and counter-rotation that a particle can experience, independent of its orbital direction relative to the central object's rotation. For $s>0$, the particle’s spin is parallel to that of the compact object, whereas for $s<0$, they are antiparallel \cite{tsupko,suzuki}.

Once the components of the spin tensor are substituted into the conserved quantities, we obtain a system of two equations with two unknowns, from which we can straightforwardly solve for $p_t$ and $p_\phi$. Under the supplementary condition employed throughout the entire work, the solutions remain valid only up to $\mathcal{O}(s)$ \cite{skoupy1,skoupy2, steinhoff,costa, barausse}. Consequently, the expression for $p_t$ in terms of the energy and the total angular momentum becomes
\begin{equation}
   p_t = \mathcal{E}_t E+\mathcal{L}_tL_z
\end{equation}
and for the $\phi$ component of the 4-momentum we obtain
\begin{equation}
    p_{\phi} = \mathcal{E}_{\phi} E+\mathcal{L}_{\phi}L_z.
\end{equation}
In order to render the expressions for the components of the 4-momentum more tractable, we expand them in Taylor series up to sixth order. The series expansions of the coefficients are provided in the Appendices.

This allows us to define a Hamiltonian that depends on the energy and the total angular momentum, which in turn enables us to directly incorporate the magnitude of the spin. The Hamiltonian is defined by
\begin{equation}
\begin{aligned}
     H&=\frac{1}{2}\left(\dot{t}p_t+X\dot{r}^2+\dot{\phi}p_{\phi}\right)= \epsilon
\end{aligned}
\end{equation}
from which we derive the new effective potential, that depends explicitly on $s$, and is expressed as follows
\begin{equation}
    \begin{aligned}
    V_\text{eff}= -\frac{2\epsilon}{X}+\frac{1}{X\mathcal{R}}\left(
    - Z p_{t}^{2} + 2W p_t p_{\phi} + V p_{\phi}^2\right)
    \end{aligned}    
\end{equation}
which corresponds to the generalization of the effective potential previously derived for spinless particles. Clearly, by setting $s=0$, we recover the previously derived result (\ref{pot}) reported by \cite{Frutos2}, where all symbols retain the same meaning as defined above. We define $\hat{s}$ as the specific spin $s/m$ ($[\hat{s}]=M$ in geometrized units). The effective potential for spinning particles is then

\begin{equation}
\begin{aligned}
V_\text{eff}&=-2\epsilon\mathcal{A}+L^2_z\left(\mathcal{B}-2J\hat{s}u^5-17Jq\hat{s} u^8\right) \\
&\quad - 2Mu^3 (L_z - aE)^2 + L_zE(7 Jqu^6 +\mathcal{D}) \\
&\quad -E^2\left(\mathcal{C}+6J\hat{s}u^3+6Jq\hat{s}u^6\right)
\end{aligned}
\end{equation} 
with 
\begin{align*}
    \mathcal{D} = 2\hat{s}\left(-u^2 + 3Mu^3 + 2Ma^2u^5 + \frac{3}{2}qu^5 + \frac{33}{4}Mqu^6\right). 
\end{align*}
To highlight the changes in comparison, we recall the definitions of the coefficients for the spinless potential presented in (\ref{pot}). We can notice how the metric term $W_q$ introduces an additional contribution to the cross-term, besides the entirely new ones coupled with the spin parameter.

\begin{figure}
    \centering
        \includegraphics[width=\linewidth]{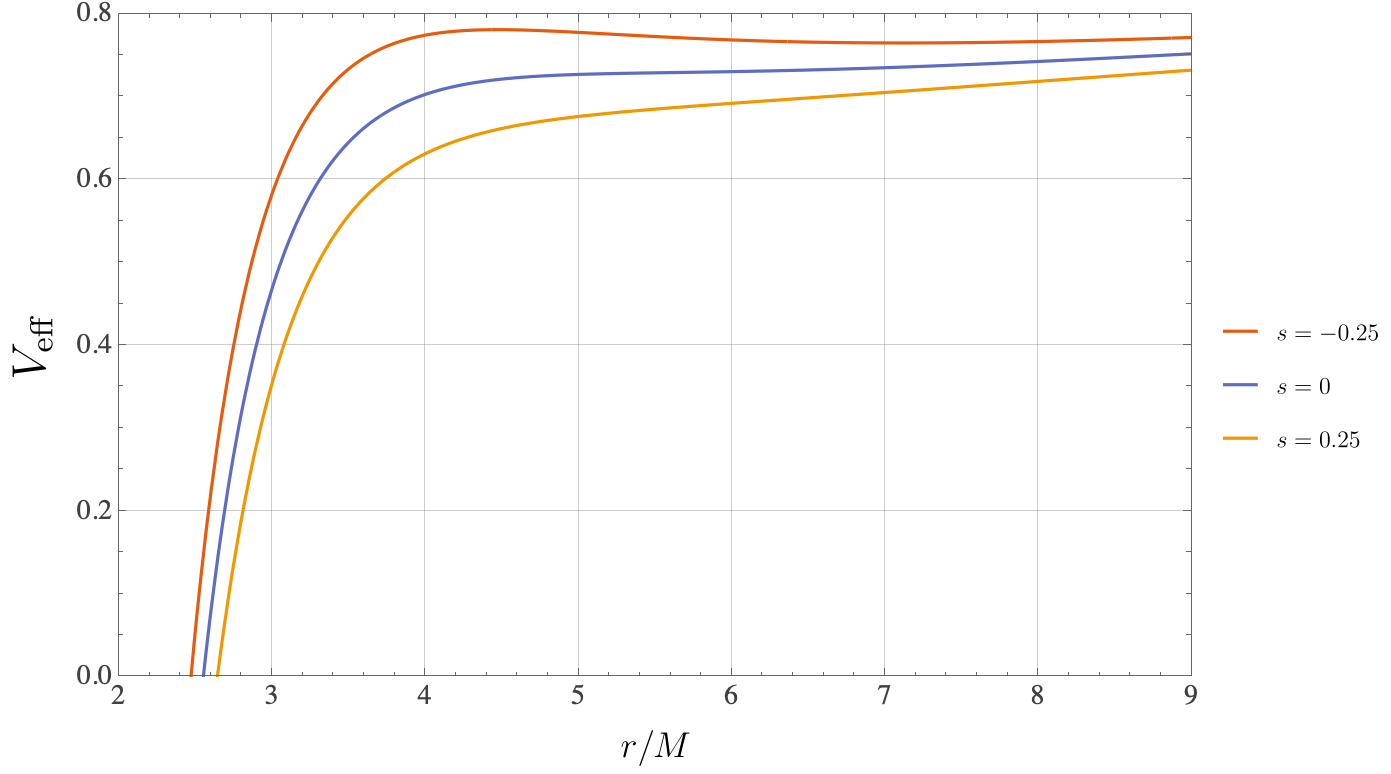}
        \caption{Effective potential $V_{\rm eff}$ for spinning test particles with $\hat{s}=-0.25,\,0,\,0.25$ (in units of $M$), for $E=1$, $L_z=4.23$, $a=M$, and $q=-2$.}
        \label{fig:vspin}
\end{figure}

Figure \ref{fig:vspin} shows this effective potential for three values of the spin parameter. The $s=0$ case reproduces the slightly different effective potential obtained previously for a spinless particle. When the particle spin is taken into account, the effective potential is modified, with the magnitude and sign of the spin determining the direction of this modification. In particular, for the parameters considered here, a positive spin parameter produces a lower effective potential than the negative-spin case, while the $s=0$ curve lies between them. This behavior reflects the coupling between the spin of the test particle and the rotation of the central body. Here, $s>0$ corresponds to a configuration in which the particle spin is aligned with the rotation of the central object, whereas $s<0$ corresponds to an anti-aligned configuration. The difference between the three curves becomes progressively smaller at larger radii, indicating that the spin-dependent interaction is strongest in the strong-field region close to the central object. This dependence on the sign of $s$ also shows that particles with the same orbital angular momentum $L_z$ can experience different effective potentials depending on their spin orientation. Consequently, when both the direction of the orbital angular momentum and the orientation of the particle spin relative to the central object's rotation are taken into account, four distinct configurations can be identified: two corresponding to each orientation of the orbital angular momentum. When the two possible orientations of the orbital angular momentum are considered together with the two possible orientations of the particle spin, four distinct configurations arise as shown in Table \ref{tab:spin_configurations}.

\begin{table}
    \begin{ruledtabular}
    \begin{tabular}{ccl}
        $L_z$ & $s$ & Configuration \\
        \hline
        $+$ & $+$ & Co-rotating, spin aligned \\
        $+$ & $-$ & Co-rotating, spin anti-aligned \\
        $-$ & $+$ & Counter-rotating, spin aligned \\
        $-$ & $-$ & Counter-rotating, spin anti-aligned
    \end{tabular}
    \end{ruledtabular}
    \caption{Classification of orbital and spin configurations according
    to the signs of $L_z$ and $s$.}
    \label{tab:spin_configurations}
\end{table}

\section{Master ISCO equation}
\label{sec:miem}

To determine the radial distances at which these orbits are located, we require the ISCO function for spinning particles. We shall therefore consider the following procedure in the most general possible manner for an effective potential. We consider the potential
\begin{equation}
     V_\text{eff}= \epsilon A(u) + B(u)L_z^2+C(u)EL_z+D(u)E^2. \label{v}
\end{equation}
The latter expression has then to be differentiated twice to find the values of $u$ where the orbit is stable
\begin{equation}
     V'_\text{eff}= \epsilon A' + B'L_z^2+C'EL_z+D'E^2. \label{vp}
\end{equation}
When differentiating the potential, $E$ and $L_z$ are held constant. Their explicit dependence on the circular orbit radius is introduced only after performing the derivatives, at which point the conditions for circularity $V_{\text{eff}} = 0$ and $V'_\text{eff}=0$ are solved simultaneously for the conserved quantities $E$ and $L_z$.

We introduce the following auxiliary quantities
\begin{equation}
    \begin{aligned}
    F_1 &= AB' - A'B, \quad F_2 = AC' - A'C, \\    
    F_3 &= AD' - A'D, \quad F_4 = B'C - BC', \\
    F_5 &= BD' - B'D, \quad F_6 = CD' - C'D.  
\end{aligned}
\end{equation}
Additionally, we define
\begin{equation}
    \begin{aligned}
    \Delta_1 &= F_2^2 - 4F_1F_3, \\ 
    \Delta_2 &= F_5^2 + F_4F_6, \\
     \Pi &= 2F_1F_5 + F_2F_4 .
    \end{aligned}
\end{equation}

Multiplying (\ref{vp}) by $A$ and subtracting $A'$ times (\ref{v}), the term proportional to $\epsilon$ is eliminated 
\begin{equation}
    F_1 L_z^2 + F_2 EL_z + F_3 E^2 = 0
\end{equation}
and for $E \neq 0$, we are able to introduce
\begin{equation}
    \Psi = \frac{L_z}{E}
\end{equation}
thus
\begin{equation}
    F_1\Psi^2 + F_2 \Psi + F_3 = 0.
\end{equation}
The discriminant of this equation is exactly $\Delta_1$. Therefore, for $F_1 \neq 0$ we have
\begin{equation}
    \Psi_{\pm}=\frac{-F_2 \pm \sqrt{\Delta_1}}{2F_1}.
\end{equation}
The solutions to the circular orbit conditions can be expressed as
\begin{equation}
E_{\pm} = \pm {\sqrt{\frac{\epsilon\,\Xi}{2\Delta_2}}},
\end{equation}
with
\begin{equation}
    \Xi = \Pi \pm F_4 \sqrt{\Delta_1}
\end{equation}
and
\begin{equation}
L_{z\pm} = E_{\pm} \Psi_{\pm}.
\end{equation}
This representation preserves the radical products of the original algebraic expressions; the branches must be selected in a consistent manner. The four distinct pairs of solutions are thus explicitly given by
\begin{equation}
    \begin{aligned}
        (E, L_z)_1 &= (E_{+}, E_{+} \Psi_{+}), \\
        (E, L_z)_2 &= (-E_{+}, -E_{+} \Psi_{+}), \\
        (E, L_z)_3 &= (E_{-}, E_{-} \Psi_{-}), \\
        (E, L_z)_4 &= (-E_{-}, -E_{-} \Psi_{-}). 
    \end{aligned}
\end{equation}
The quadratic combination necessary for evaluating the stability condition is
\begin{equation}
\label{quadratic}
E^2 = \frac{\epsilon\,\Xi}{2\Delta_2},   
\end{equation}
where we evidently have $EL_z = \Psi_{\pm} E^2$ and $L_z^2 = \Psi_{\pm}^2 E^2.$ These quadratic forms are independent of the sign choice. Consequently, the four solutions yield only two physically distinct expressions for the stability condition. Alternatively, using the first circular orbit condition, the squared energy can be written as
\begin{equation}
E^2 = -\frac{\epsilon A}{B \Psi_{\pm}^2 + C \Psi_{\pm} + D}, \label{eq:En2_alt}
\end{equation}
as long as the denominator does not vanish.

The second derivative of the effective potential with respect to $u$, holding $E$ and $L_z$ constant, is defined as
\begin{equation}
V''_\text{eff} = \epsilon A'' + B'' L_z^2 + C'' E L_z + D'' E^2.
\end{equation}
Substituting the quadratic combinations from (\ref{quadratic}) yields
\begin{equation}
V''_{\text{eff}\pm} = \epsilon \left[ A'' + \frac{\Xi}{2\Delta_2} \left( B'' \Psi_{\pm}^2 + C'' \Psi_{\pm} + D'' \right) \right].
\end{equation}
We now introduce another auxiliary quantity defined as
\begin{align}
Q_{\pm} = \pm\sqrt{\Delta_1} - F_2 \implies
\Psi_{\pm} = \frac{Q_{\pm}}{2F_1}
\end{align}
and the expression for $V''_{\text{eff}\pm}$ then becomes
\begin{equation}
V''_{\text{eff}\pm} = \epsilon \biggl[ A'' + \frac{\Xi\, \Phi}{8\Delta_2 F_1^2} \biggr].
\end{equation}
where
\begin{equation*}
    \Phi = B'' Q_{\pm}^2 + 2C'' F_1Q_{\pm} + 4D'' F_1^2 \, .
\end{equation*}
For $\epsilon \neq 0$, $\Delta_2 \neq 0$, and $F_1 \neq 0$, the marginal stability condition $V''_{\text{eff}\pm} = 0$ is equivalent to finding the roots of
\begin{equation}
\mathcal{P}_{\pm} = 8\Delta_2 F_1^2 A'' + \Xi \, \Phi = 0.
\end{equation}
An alternative equivalent form derived from Eq.~\eqref{eq:En2_alt} is
\begin{equation}
\mathcal{P}_{\pm} = A'' - A \frac{B'' \Psi_{\pm}^2 + C'' \Psi_{\pm} + D''}{B \Psi_{\pm}^2 + C \Psi_{\pm} + D},
\end{equation}
which holds provided that the denominator remains non-zero. 

To express the ISCO function in a completely compact form, we can introduce new auxiliary quantities defined in terms of the components of the effective potential and their derivatives. To further simplify the ISCO function, we introduce two new auxiliary quantities defined as
\begin{equation}
\begin{aligned}
F_7 &=B''\left(F_2^2+\Delta_1\right)-2C''F_1F_2+4D''F_1^2,\\
F_8 &=2B''F_2-2C''F_1.
\end{aligned}
\label{eq:HK_ISCO}
\end{equation}
Using these definitions, the quantities $\Delta_3$ and $\Delta_4$ can be written as
\begin{equation}
\begin{aligned}
\Delta_3 &=-F_4F_7+\Pi F_8,\\
\Delta_4 &= 8\Delta_2F_1^2A''+\Pi F_7-F_4\Delta_1F_8.
\end{aligned}
\label{eq:Delta34}
\end{equation}
The ISCO function then takes the compact form
\begin{equation}
\mathcal{P}_{\mathrm{ISCO}}=\pm\sqrt{\Delta_1}\,\Delta_3+\Delta_4,
\label{eq:P_ISCO}
\end{equation}
where $\pm$ specifies the algebraic branch. We call this the master ISCO equation. The ISCO radius is determined by solving $\mathcal{P}_{\mathrm{ISCO}}=0$ for the radial coordinate, subject to the physical constraints of the spacetime and the particle configuration. Again, all series expansions of the coefficients and auxiliary variables employed in this procedure are provided in the Appendices. Specifically, Appendix \ref{app:series} contains the expanded and explicit forms of all variables given here.

To determine the energy $E$ and angular momentum $L_z$ for spinning test particles in the ISCO, the analytical procedure proceeds through the following steps. First we evaluate the effective potential coefficients and their derivatives for the desired configuration to compute $F_1, \dots, F_6$, along with $\Delta_1$, $\Delta_2$, and $\Pi$. With this we are now able to compute the energy, taking the positive branch $E > 0$, and the angular momentum, where the sign of $L_z$ is determined by the sign of $\Psi$ (since $E > 0$). Subsequently, a perturbative expansion in the particle's spin $s$ is applied to $E$ and $L_z$ around the non-spinning ISCO radius $r_0$ using the linear correction $\delta r$ from $r_{\rm ISCO} = r_0 + \delta r$, yielding closed-form analytical expressions up to first order in spin. We find this by solving the ISCO equation $\mathcal{P}_{\rm ISCO} = \pm\sqrt{\Delta_1}\,\Delta_3 + \Delta_4 = 0$ for the specific spacetime to obtain the ISCO root $r_{\rm ISCO}$. 
\section{Limiting Cases}
\label{sec:lit}

We now examine our results in the limiting cases of the KLMQ metric for the allowed values of $r$. This allows us to directly determine the ISCO function for each case and its corresponding characteristic parameters. Every ISCO function is written as $\mathcal P=0$ with global factors removed. $E^2$ and $L_z^2$ are circular-orbit values at any $r$ and are evaluated at the root of $\mathcal P$ to obtain the ISCO values. 

\subsection{Schwarzschild}
We then begin with the simplest case contained within the metric. For the Schwarzschild spacetime \cite{schwarzschild} ($a = q = 0$), the ISCO function takes the following form
\begin{equation}
    \mathcal{P}_{\mathrm{S}}= (r-3M)(r-6M) \pm 3\hat{s}(r-2M)\sqrt{\frac{M}{r}},
\end{equation}
where we can explicitly identify within the function both the ISCO at $6M$ and the photon sphere at $3M$, thereby confirming the correct physical behavior in the Schwarzschild limit for spinless particles. From the method, we can then obtain compact expressions for the energy and angular momentum including linear-order spin corrections, which are given by
\begin{gather}
    E^2_{\mathrm{S}} = \frac{\left(r - 2 M\right)^{2}}{r \left(r - 3 M\right)} \mp \hat{s}\frac{\left(r - 2 M\right)}{\left(r - 3 M\right)^{2}}\sqrt{\frac{M^3}{r^3}}, \\
    L^2_{z\mathrm{S}} = \frac{M r^{2}}{\left(r - 3 M\right)} \pm \hat{s}\frac{\sqrt{Mr}\left(r - 2 M\right) \left(2 r - 9 M\right)}{\left(r - 3 M\right)^{2}}.
\end{gather}

We now consider a small-spin perturbation around the ISCO value by writing
\begin{equation}
r=6M+\delta r,\quad\delta r=\mathcal{O}(s).
\end{equation}
Consequently,
\begin{equation}
r-6M=\delta r,\quad r-3M=3M+\delta r,
\end{equation}
and the first term of the function becomes
\begin{align*}
(r-3M)(r-6M)&=(3M+\delta r)\delta r\\
&=3M\delta r+(\delta r)^2.
\end{align*}
Since $\delta r=\mathcal{O}(s)$, the second term is of order
$\mathcal{O}(s^2)$ and can be neglected when retaining only terms
linear in the spin. Thus,
\begin{equation}
(r-3M)(r-6M)\simeq3M\delta r. \label{rhs}
\end{equation}
The second term in $\mathcal{P}_{\mathrm{S}}$ already contains an explicit factor of $s$. Therefore, corrections arising from replacing $r=6M+\delta r$ inside the remaining factors would be of order $\mathcal{O}(s^2)$. To first order in $s$, we can consequently evaluate these factors at the zeroth-order value $r=6M$
\begin{align}
3\hat{s}(r-2M)\sqrt{\frac{M}{r}}&\simeq  2\sqrt{6}\,\hat{s}M.
\label{lhs}
\end{align}
Substituting \eqref{rhs} and \eqref{lhs}, we obtain, to linear order in the spin,
\begin{equation}
3M\delta r \pm 2\sqrt{6}\,\hat{s}M =0.
\end{equation}
Solving for the radial perturbation gives
\begin{equation}
\delta r = \mp \frac{2\sqrt{6}}{3}\hat{s} = \mp\, 2\sqrt{\frac{2}{3}}\hat{s}.
\end{equation}
Hence, the ISCO radius to first order in the spin is
\begin{equation}
r_{\rm ISCO} \simeq 6M \mp 2\sqrt{\frac{2}{3}}\hat{s}.
\end{equation}
This result provides a consistency check of the ISCO function in the
Schwarzschild limit, just as in \cite{tsupko,oleg}. In the following, $E_{\rm ISCO}$ and $L_{z\,{\rm ISCO}}$ denote the energy and the total angular momentum in the innermost stable circular orbit. Again, $s>0$ corresponds to a particle spin parallel to the rotation of the central body and only terms up to first order in the small parameters are retained. Applying the perturbative expansion to $E$ and $L_z$ ultimately yields
\begin{equation}
\begin{aligned}
  E_{\rm ISCO} &= \frac{2\sqrt{2}}{3} \mp \frac{1}{36\sqrt{3}}\,\frac{\hat{s}}{M}, \\
  L_{z\,{\rm ISCO}} &= \pm\, 2\sqrt{3}\,M + \frac{\sqrt{2}}{3}\,\hat{s},
\end{aligned} 
\end{equation}
which are also in consistent agreement with those reported in \cite{tsupko,oleg}.

\subsection{Kerr}
For the full Kerr spacetime \cite{kerr}, incorporating the exact dependence on the rotation parameter ($a\neq0$ and $q=0$), the ISCO function takes the form
\begin{equation}
\begin{aligned}
        \mathcal{P}_{\mathrm{K}}&= r\,\big[\,r\,(r-2M)^2-Ma^2\big]P_{\rm K}\\
        &\quad-M\hat{s}\Big(a\,r\,\mathcal{S}_1\mp3\sqrt{Mr}\Delta\,\mathcal{S}_2\Big),
\end{aligned}
\end{equation}
where
\begin{equation*}
\begin{aligned}
    P_{\rm K}&=Mr^5-9M^2r^4+3(6M^3-Ma^2)r^3\\
    &\quad-7M^2a^2r^2\pm6Mar\sqrt{Mr}\,\Delta,\\
    \mathcal{S}_1&=3r^6+31Mr^5+\left(15a^2-316M^2\right)r^4\\
    &\quad+48M\left(17M^2-a^2\right)r^3-M^2\left(648M^2+205a^2\right)r^2\\
    &\quad+Ma^2\left(466M^2-33a^2\right)r-81M^2a^4,\\
    \mathcal{S}_2 &=r^5-4Mr^4+(13a^2+4M^2)r^3-73Ma^2r^2\\
    &\quad+84M^2a^2r-25Ma^4.
\end{aligned}
\end{equation*}
 The compact expressions for the energy and angular momentum including linear-order spin corrections for the Kerr background are the following
\begin{widetext}
\begin{gather}
        E^2_{\mathrm{K}} = \frac{\left(r^{2} - 2 M r \pm a \sqrt{Mr}\right)^{2}}{r^{2} \left(r^{2} - 3 M r \pm 2a \sqrt{Mr}\right)}\mp \hat{s}\frac{M \left(\sqrt{Mr} \mp  a\right) \left(r^{2} - 2 M r \pm a \sqrt{Mr}\right) \left(r^{2} \mp 4 a \sqrt{Mr} + 3 a^{2}\right)}{r^{3} \left(r^{2} - 3 M r \pm 2a \sqrt{Mr}\right)^{2}},\\
        L^2_{z\mathrm{K}} = \frac{M \left(r^{2} \mp 2 a \sqrt{Mr} + a^{2}\right)^{2}}{r \left(r^{2} - 3 M r \pm 2 a \sqrt{Mr}\right)} \pm \hat{s}\frac{\sqrt{M} \left(r^{2} \mp 2 a \sqrt{Mr} + a^{2}\right) \mathcal N_{\text{K}}}{\sqrt{r^5}\left(r^{2} - 3 M r \pm 2 a \sqrt{Mr}\right)^{2}},
\end{gather}
where
\begin{align*}
\mathcal N_{\text{K}} &= 2 r^{5} - 13 M r^{4} \pm 9 a r^{3} \sqrt{Mr} + 18 M^{2} r^{3} \mp 21 a M r^{2} \sqrt{Mr} + 2 a^{2} M r^{2} \pm 3 a^{3} r \sqrt{Mr}  + 4 a^{2} M^{2} r \mp 7 a^{3} M \sqrt{Mr} + 3 a^{4} M.
\end{align*}
\end{widetext}

Next, we perform a perturbative expansion about the counter-rotating ISCO radius in an extremal Kerr background, $r_0 = 9M$, corresponding to $a = M$ \cite{bardeen}. Our analysis is restricted to terms linear in the particle's spin $s$, while discarding higher-order spin contributions. We write
\begin{equation}
r=9M+\delta r,
\quad
\delta r=\mathcal{O}(s).
\end{equation}
After performing a similar procedure, we find
\begin{equation}
r_{\mathrm{ISCO}} \simeq 9M+\frac{16}{9}\hat{s}
\end{equation}
This result gives the leading-order correction to the retrograde ISCO radius of an extreme Kerr black hole due to the spin of the orbiting particle, just as determined by \cite{oleg,tsupko}. The sign of the correction is determined by the adopted convention for the particle's spin orientation relative to the orbital angular momentum.

For the co-rotating ISCO in the extreme Kerr limit, we attempted a perturbative expansion around the geodesic ISCO radius by setting
\begin{equation}
r=M+\delta r, \quad \delta r=\mathcal{O}(s).
\end{equation}
However, this procedure does not yield a linear correction to $r_{\mathrm{ISCO}}$ in the particle's spin. This behavior is associated with the degeneracy of the co-rotating ISCO in the extreme Kerr limit, which makes the standard linear expansion inconclusive for this particular case. This aligns with previous findings in the literature. Specifically, as noted in \cite{takahiro} on the basis of numerical computations, in the extreme Kerr background for the parallel spin configuration, the magnitude of the test body's spin does not affect the radius of the innermost stable circular orbit, which remains fixed at $r_{\text{ISCO}} = M$. Around the geodesic ISCO radius $r_0=9M$ we obtain for the energy $E$ and the angular momentum $L_z$
\begin{equation}
 \begin{aligned}
  E_{\rm ISCO} &= \frac{5\sqrt{3}}{9} + \frac{\sqrt{3}}{243}\,\frac{\hat{s}}{M}, \\
  L_{z\,{\rm ISCO}} &= -\frac{22\sqrt{3}}{9}\,M + \frac{82\sqrt{3}}{243}\,\hat{s}.
\end{aligned}   
\end{equation}
For the co-rotating extremal case, $r_0=M$, the linear expansion in $s$ is
degenerate and no linear correction is obtained.

\subsection{Lense-Thirring}
Finally, the Lense-Thirring spacetime \cite{lensethirring,mashhoon1984}, accounting for frame-dragging effects in the slow-rotation limit ($q=0$ and keeping terms linear in $a$), the ISCO function is given by
\begin{equation}
\begin{aligned}
\mathcal{P}_{\mathrm{LT}}&= r(r-2M)^2P_{\rm LT}\\
&\quad-M\hat{s}\Big[a\,\mathcal{S}_1^{l}\mp3\sqrt{M}r^{7/2}(r-2M)^3\Big],
\end{aligned}
\end{equation}
where
\begin{equation*}
    \begin{aligned}
       P_{\rm LT} &= Mr^3(r-3M)(r-6M) \pm 6Mar^2\sqrt{Mr}\,(r-2M),\\
       \mathcal{S}_1^{l} &= 3r^6+31Mr^5-316M^2r^4+816M^3r^3-648M^4r^2.
    \end{aligned}
\end{equation*}

For the Lense-Thirring spacetime, the compact formulas for the energy and angular momentum, incorporating linear-order spin corrections, read as follows
\begin{widetext}
\begin{gather}
        E^2_{\mathrm{LT}} = \frac{\left(r - 2 M\right) \left(r^{3} - 5 M r^{2} + 6 M^{2} r \mp 2 a M \sqrt{Mr}\right)}{r^{2} \left(r - 3 M\right)^{2}} \mp \hat{s}\frac{M \mathcal N_{\text{LT},e}}{r^{3} \left(r - 3 M\right)^{3}},\\
        L^2_{z\mathrm{LT}} = \frac{M \left(r^{3} - 3 M r^{2} \mp 6 a r \sqrt{Mr} \pm 12a M \sqrt{Mr}\right)}{\left(r - 3 M\right)^{2}} \pm \hat{s}\frac{\sqrt{M} \mathcal N_{\text{LT},\ell}}{\sqrt{r^3}\left(r - 3 M\right)^{3}},
\end{gather} 
where 
\begin{gather*}
\mathcal N_{\text{LT},e} = r^{3} \sqrt{Mr} \mp a r^{3} - 5 M r^{2} \sqrt{Mr} \mp 2 a M r^{2} + 6 M^{2} r \sqrt{Mr} \pm 19 a M^{2} r \mp 24 a M^{3},\\
\mathcal N_{\text{LT},\ell} = 2 r^{5} - 19 M r^{4} \mp 3  a r^{3} \sqrt{Mr} + 57 M^{2} r^{3}  \pm 42 a M r^{2} \sqrt{Mr} - 54 M^{3} r^{2} \mp 123 a M^{2} r \sqrt{Mr}  \pm 108 a M^{3} \sqrt{Mr}. 
\end{gather*}
\end{widetext}

We now consider a perturbative expansion around the Schwarzschild ISCO radius, retaining terms linear in the central body's rotation parameter $a$, linear in the particle's spin $s$, and bilinear in $as$. We write
\begin{equation}
r=6M+\delta r,\quad \delta r=\delta r_a+\delta r_s+\delta r_{as},
\end{equation}
where $\delta r_a=\mathcal{O}(a)$, $\delta r_s=\mathcal{O}(s)$, and $\delta r_{as}=\mathcal{O}(as)$. Terms proportional to $a^2$ and $s^2$ will be neglected. Since we retain the mixed contribution proportional to $as$, the quadratic term cannot be discarded entirely. Instead, we keep only the cross term between the first-order corrections:
\begin{equation}
(\delta r)^2
\simeq 2\delta r_a\delta r_s,
\end{equation}
where terms proportional to $a^2$, $s^2$, or higher perturbative orders have been omitted. After doing a similar procedure, we obtain
\begin{equation}
r_{\mathrm{ISCO}}\simeq6M\mp\,4\sqrt{\frac{2}{3}}\,a\mp\,2\sqrt{\frac{2}{3}}\hat{s}+\frac{2}{9}\frac{a\hat{s}}{M}.
\end{equation}
The terms linear in $a$ and $s$ reproduce the slowly rotating Kerr and Schwarzschild spinning-particle limits, respectively. The mixed term describes the coupling between the rotation of the central body and the particle's spin. Its coefficient can be compared with the corresponding result in \cite{tsupko,oleg}. Keeping terms linear in $a$, linear in $s$, and the bilinear contribution $\propto as$ for the perturbative expansion applied to $E$ and $L_z$ gives us finally 
\begin{equation}
\begin{aligned}
  E_{\rm ISCO} &= E_{\text{ISCO,S}}\mp \frac{1}{18\sqrt{3}}\,\frac{a}{M}- \frac{\sqrt{2}}{81}\,\frac{a\hat{s}}{M^{2}},\\
  L_{z\,{\rm ISCO}} &= L_{z\,\text{ISCO,S}}- \frac{2\sqrt{2}}{3}\,a \mp \frac{11}{36\sqrt{3}}\,\frac{a\hat{s}}{M}.
\end{aligned}  
\end{equation}
The terms linear in $s$ and $a$ reproduce the Schwarzschild spinning-particle
result and the slowly rotating Kerr geodesic result, respectively, while the
last term in each expression describes the coupling between the rotation of
the central body and the particle's spin.

A summary of the key results for all limiting cases is presented in Table~\ref{tab:limiting_cases}, where the Schwarzschild, Lense-Thirring, and full Kerr limits are explicitly compared. We can observe how, for the Lense-Thirring case, the Schwarzschild results emerge explicitly, accompanied by additional corrections that depend on the specific parameter of the spacetime.

\begin{table*}
\begin{ruledtabular}
\begin{tabular}{cccc}
Limiting Case & $r_{\text{ISCO}}$ & $E_{\text{ISCO}}$ & $L_{z\,\text{ISCO}}$ \\[0.5ex]
\hline
\rule{0pt}{4ex}%
S ($a = 0, \, q = 0$) &  
$6M \mp 2\sqrt{\dfrac{2}{3}}\hat{s}$ & $\dfrac{2\sqrt{2}}{3}\mp\dfrac{1}{36\sqrt{3}}\dfrac{\hat{s}}{M}$ & $\pm2\sqrt{3}M+\dfrac{\sqrt{2}}{3}\hat{s}$\\[2.ex]
\hline
\rule{0pt}{4ex}%
LT ($a \ll M, \, q = 0$) &  $r_{\text{ISCO,S}} \mp\,4\sqrt{\dfrac{2}{3}}\,a+\dfrac{2}{9}\dfrac{a\hat{s}}{M}$ & $E_{\text{ISCO,S}} \mp \dfrac{1}{18\sqrt{3}} \dfrac{a}{M} - \dfrac{\sqrt{2}}{81} \dfrac{a\hat{s}}{M^2}$ & $L_{z\,\text{ISCO,S}} - \dfrac{2\sqrt{2}}{3} a \mp \dfrac{11}{36\sqrt{3}} \dfrac{a\hat{s}}{M}$ \\[2.5ex]
\hline
\rule{0pt}{4ex}%
K ($a = M, \, q = 0$) &  
$9M+\dfrac{16}{9}\hat{s}$  & $\dfrac{5\sqrt{3}}{9} + \dfrac{\sqrt{3}}{243}\dfrac{\hat{s}}{M}$ & $-\dfrac{22\sqrt{3}}{9}M + \dfrac{82\sqrt{3}}{243}\hat{s}$\\[1ex]
\end{tabular}
\end{ruledtabular}
\caption{\label{tab:limiting_cases}Limiting spacetimes encompassed within the KLMQ, showing their corresponding radii ($r_{\text{ISCO}}$), energies ($E_{\text{ISCO}}$), and angular momenta ($L_{z\,\text{ISCO}}$).}
\end{table*}
\section{ISCO relations}\label{sec:iscorel}

Using the derived expressions for both energy $E$ and angular momentum $L_z$, we can extract further relations for the ISCO. The impact parameter $b$ of a particle with energy $E$ and angular momentum $L_z$ is computed from the conserved quantities as
\begin{equation}
    b = \frac{L_z}{E}.
\end{equation}
At the ISCO, $b_{\text{ISCO}} = L_{z,\text{ISCO}}/E_{\text{ISCO}}$. However, for simplicity, we compute $b^2$ instead. Since $\hat{s}$ is small, we can perform a first-order Taylor expansion of the ratio to obtain compact expressions with the geodesic value plus the leading spin correction.

For the Schwarzschild spacetime we have that 
\begin{equation}
    b^2_{\text{S}}=\frac{M r^3}{(r-2M)^2}\left[1\pm\,\hat{s}\,\frac{(2r-3M)(r-4M)}{\sqrt M\,r^{3/2}\,(r-2M)}\right].
\end{equation}
Evaluating this relation at the ISCO yields
\begin{equation}
    b^{2,\text{S}}_{\text{ISCO}} = \frac{27}{2}M^{2} \pm \frac{27\sqrt{6}}{16}\,M\hat{s}.
\end{equation}
A completely analogous procedure can be followed to derive the corresponding quantity in the Kerr spacetime. The compact expression for the impact parameter is given by
\begin{equation}
    b^2_{\text{K}}=\frac{M r\,\big(r^2\mp2 a\sqrt{Mr}+a^2\big)^2}{\big(r^2-2Mr\pm a\sqrt{Mr}\big)^2}\left(1\pm\,\hat{s}\,\frac{\kappa_1}{\kappa_2}\right),
\end{equation}
where
\begin{align*}
    \kappa_1&=r^2(2r-3M)(r-4M) \\
    &\quad\pm a\sqrt{Mr}\,\big(7r^2-10Mr+3a^2\big)-3a^2Mr,\\
    \kappa_2&=\sqrt{Mr}\,\big(r^2\mp2 a\sqrt{Mr}+a^2\big)\big(r^2-2Mr\pm a\sqrt{Mr}\big).
\end{align*}
If we now consider the counter-rotating case and evaluate it at $r_0 = 9M$, we obtain
\begin{equation}
    b^{2,\text{K}}_{\text{ISCO}} = \frac{484}{25}M^{2}-\frac{704}{125}\,M\hat{s}.
\end{equation}
Finally, for the Lense-Thirring metric, the impact parameter reads
\begin{equation}
b^2_{\rm LT}=\frac{M r^{3}}{(r-2M)^{2}}\left[1\mp2a\,\frac{\sqrt{Mr}\,(3r-4M)}{r^{2}(r-2M)}\pm\,\hat{s}\,\kappa_{\rm S}\right],
\end{equation}
where
\begin{equation*}
\kappa_{\rm S} = \frac{(2r-3M)(r-4M)}{\sqrt M\,r^{3/2}(r-2M)}
\end{equation*}
just as in Schwarzschild. Evaluating these expressions at $r = 6M$ yields the corresponding Lense-Thirring values
\begin{equation}
    b^{2,\text{LT}}_{\rm ISCO} = \frac{27}{2}M^{2} \mp \frac{21\sqrt{6}}{8}\,M a \pm \frac{27\sqrt{6}}{16}\,M\hat{s}.
\end{equation}

\section{Conclusions}\label{sec:conclu}

We investigated the dynamics of test particles in the ISCO for an approximate Kerr-like metric with a mass quadrupole moment. For both spinless and spinning particles, several noteworthy behaviors were identified. From the analysis of spinless particles, it is already known that two types of motion arise when orbiting a central body. This phenomenon significantly alters the radial distance at which the particle can orbit. In addition, numerical computations could allow us to determine graphically how the quadrupole moment affects this distance, either pushing the particle inward or outward depending on the magnitude of the angular momentum.

For spinning particles, the quadrupole moment enters the conserved quantities and the spin tensor in specific ways. From the conserved quantities, we observe how the quadrupole influences specific components of the spin tensor ---most notably the $tr$ component in the expression for the energy and the $r\phi$ component in the total angular momentum. Moreover, the solutions corresponding to the Schwarzschild and Kerr limits are readily recovered. Once the supplementary conditions are imposed and the explicit form of the spin–tensor components is determined, the direct influence of the quadrupole becomes evident through the function that introduces it into the metric up to second order. More generally, any additional parameter introduced perturbatively into the metric (e.g., a mass quadrupole or a magnetic dipole moment) enters the spin tensor and the conserved quantities and therefore modifies the ISCO parameters. 

In addition to the typical co-rotating and counter-rotating motions experienced by a particle orbiting a rotating central body, further adjustments arise that can affect the radial distance of the orbit. This occurs because the magnitude of the particle’s spin also contributes to the alignment between angular momenta. For each orbital direction, there exist two possible spin orientations, resulting in a total of four distinct orbital radii when the particle’s spin is taken into account.

The generalization of the ISCO equation for spinning particles allows for a more realistic and physically meaningful treatment of the problem of a particle orbiting within an accretion disk around a massive central object. If the ISCO radius of a compact object, such as a neutron star, could be inferred observationally, the ISCO polynomial would provide an indirect estimate of its quadrupole moment. 

A novel procedure utilizing the effective potential has been introduced, allowing for the derivation of a general ISCO equation valid for both spinning and non-spinning test particles, as well as for any additional parameters characterizing the background metric. Within this framework, we have obtained compact expressions for the most widely studied limiting cases, verifying that the expected spin corrections to the various physical quantities are accurately recovered. Furthermore, $\mathcal{P}_\text{ISCO}$, the energy, and the angular momentum are now cast in simplified forms that facilitate mathematical treatment, enabling us to derive new relations for key physical quantities at the ISCO.


\appendix 

\vspace*{1mm}
\begin{center}
    {\large \textbf{Appendices}}
\end{center}
\vspace{1mm}

In these appendices, all results are derived from the general Lewis metric (\ref{lewis}), without specifying the exact form of the five metric functions. The main focus is on understanding the general behavior of the metric in spherical-like coordinates, independent of the particular definition of the function's form. Later, these functions can be specified to correspond to known metrics such as the Schwarzschild or Kerr solution, and compute the resulting expressions explicitly. Later, we write out all the series expansions employed in obtaining the ISCO function, ranging from the effective potentials to the various auxiliary quantities, which collectively govern the radial stability of the spinning test particle.

\section{Contravariant metric}
From (\ref{lewis}), it follows that the contravariant metric is
\begin{equation*}
g^{\mu\nu}=
\begin{pmatrix}
-\dfrac{Z}{\mathcal{R}} & 0 & 0 & \dfrac{W}{\mathcal{R}} \\
0 & \dfrac{1}{X} & 0 & 0 \\
0 & 0 & \dfrac{1}{Y} & 0 \\
\dfrac{W}{\mathcal{R}} & 0 & 0 & \dfrac{V}{\mathcal{R}}
\end{pmatrix} 
\end{equation*}
where $\mathcal{R}=W^2+VZ$.

\section{Killing vectors and conserved quantities} \label{sec:appe}
The Killing vectors for the general Lewis metric are the following. The time-like Killing vector is
\begin{equation*}
    \xi^{(t)}_\mu = V\delta^t_\mu + W\delta^{\phi}_\mu 
\end{equation*}
and the space-like axial Killing vector is given by
\begin{equation*}
    \xi^{(\phi)}_\mu = W\delta^t_\mu + Z\delta^{\phi}_\mu \,.
\end{equation*}

From ($\ref{quan}$), the conserved quantities associated with each symmetry are defined as follows. We have then that the general expression for the energy is 
\begin{equation*}
    \begin{aligned}
    E &= \frac{1}{2}\big( S^{\theta\phi}\partial_{\theta}W + S^{r\phi}\partial_{r}W + S^{t\theta}\partial_{\theta}V + S^{tr}\partial_{r}V \big)  -p_t   
    \end{aligned}
\end{equation*}
and the total angular momentum in the $z$-direction is
\begin{equation*}
    \begin{aligned}
        L_z &= \frac{1}{2}\big( S^{\theta\phi}\partial_{\theta}Z + S^{r\phi}\partial_{r}Z -S^{t\theta}\partial_{\theta}W - S^{tr}\partial_{r}W \big) +p_{\phi} \,.
    \end{aligned}
\end{equation*}

\onecolumngrid 

\section{Christoffel symbols}
The Christoffel symbols corresponding to the metric used are as follows
\begin{equation*}
\begin{aligned}
    \Gamma^t_{tt} &= 0\,,  &\quad \Gamma^r_{tt}  &= \frac{\partial_r V}{2 X}\,, &\quad \Gamma^{\theta}_{tt} &= \frac{\partial_{\theta}V}{2 Y}\,, &\quad \Gamma^{\phi}_{tt} &= 0\,, \\[2ex]
    \Gamma^t_{tr} &= \frac{Z\partial_r V + W\partial_r W}{2 \mathcal{R}}\,, & \Gamma^r_{tr} &= 0\,, & \Gamma^{\theta}_{tr} &= 0\,, & \Gamma^{\phi}_{tr} &= \frac{V\partial_rW- W\partial_rV}{2 \mathcal{R}}\,, \\[2ex]
    \Gamma^t_{t\theta} &= \frac{Z\partial_{\theta}V +W\partial_{\theta}W}{2 \mathcal{R}}\,, & \Gamma^r_{t\theta} &= 0\,, & \Gamma^{\theta}_{t\theta} &= 0\,, & \Gamma^{\phi}_{t\theta} &= \frac{V\partial_{\theta}W-W\partial_{\theta}V }{2 \mathcal{R}}\,, \\[2ex]
    \Gamma^t_{t\phi} &= 0\,, & \Gamma^r_{t\phi} &= -\frac{\partial_r W}{2 X}\,, & \Gamma^{\theta}_{t\phi} &= -\frac{\partial_{\theta}W}{2 Y}\,, & \Gamma^{\phi}_{t\phi} &= 0\,, \\[2ex]
    \Gamma^t_{rr} &= 0\,, & \Gamma^r_{rr} &= \frac{\partial_r X}{2 X}\,, & \Gamma^{\theta}_{rr} &= -\frac{\partial_{\theta}X}{2 Y}\,, & \Gamma^{\phi}_{rr} &= 0\,, \\[2ex]
    \Gamma^t_{r\theta} &= 0\,, & \Gamma^r_{r\theta} &= \frac{\partial_{\theta}X}{2 X}\,, & \Gamma^{\theta}_{r\theta} &= \frac{\partial_r Y}{2 Y}\,, & \Gamma^{\phi}_{r\theta} &= 0\,, \\[2ex]
    \Gamma^t_{r\phi} &= \frac{W\partial_{r}Z-Z\partial_{r}W }{2 \mathcal{R}}\,, & \Gamma^r_{r\phi} &= 0\,, & \Gamma^{\theta}_{r\phi} &= 0\,, & \Gamma^{\phi}_{r\phi} &= \frac{W\partial_{r}W +V\partial_{r}Z}{2 \mathcal{R}}\,,\\[2ex]
    \Gamma^t_{\theta\theta} &= 0\,, & \Gamma^r_{\theta\theta} &= -\frac{\partial_r Y}{2 X}\,, & \Gamma^{\theta}_{\theta\theta} &= \frac{\partial_{\theta}Y}{2 Y}\,, & \Gamma^{\phi}_{\theta\theta} &= 0\,, \\[2ex]
    \Gamma^t_{\theta\phi} &= \frac{W\partial_{\theta} Z-Z\partial_{\theta} W }{2 \mathcal{R}}\,, & \Gamma^r_{\theta\phi} &= 0\,, & \Gamma^{\theta}_{\theta\phi} &= 0\,, & \Gamma^{\phi}_{\theta\phi} &= \frac{W\partial_{\theta} W + V\partial_{\theta} Z}{2 \mathcal{R}}\,,\\[2ex]
    \Gamma^t_{\phi\phi} &= 0\,, & \Gamma^r_{\phi\phi} &= -\frac{ \partial_r Z}{2 X}\,, & \Gamma^{\theta}_{\phi\phi} &= -\frac{ \partial_{\theta} Z}{2 Y}\,, & \Gamma^{\phi}_{\phi\phi} &= 0\,.
\end{aligned}
\end{equation*}

\section{Riemann tensor}

 The non-zero independent components of the Riemann curvature tensor are given by the following
 \begin{align*}
R_{trtr} &= \frac{\partial_{\theta}V\partial_{\theta}X}{4Y}
          - \frac{\partial_{r}V\partial_{r}X}{4X}
          + \frac{V(\partial_rW)^2}{4\mathcal{R}}
          - \frac{W\partial_{r}V\partial_{r}W}{2\mathcal{R}} - \frac{Z(\partial_rV)^2}{4\mathcal{R}}
          + \frac{\partial_r\partial_rV}{2}\,, \\[2ex]
R_{trt\theta} &= \frac{V\partial_rW\partial_{\theta}W}{4\mathcal{R}}
          - \frac{Z\partial_rV\partial_{\theta}V}{4\mathcal{R}}
          - \frac{W}{4\mathcal{R}}\left(\partial_rV\partial_{\theta}W
          + \partial_{\theta}V\partial_rW\right) \\
         &\quad - \frac{\partial_rV\partial_{\theta}X}{4X}
          - \frac{\partial_{\theta}V\partial_rY}{4Y}
          + \frac{\partial_r\partial_{\theta}V}{2}\,, \\[2ex]
R_{trr\phi} &= \frac{\partial_{\theta}W\partial_{\theta}X}{4Y}
          - \frac{\partial_{r}W\partial_{r}X}{4X}
          + \frac{\partial_rZ}{4\mathcal{R}}\left(W\partial_rV-V\partial_rW\right) \\
         &\quad - \frac{\partial_rW}{4\mathcal{R}}\left(W\partial_rW+Z\partial_rV\right)
          + \frac{\partial_r\partial_rW}{2}\,, \\[2ex]
R_{tr\theta\phi} &= -\frac{\partial_{r}W\partial_{\theta}X}{4X}
          - \frac{\partial_{\theta}W\partial_{r}Y}{4Y}
          + \frac{\partial_rZ}{4\mathcal{R}}\left(W\partial_{\theta}V-V\partial_{\theta}W\right) \\
         &\quad - \frac{\partial_{\theta}W}{4\mathcal{R}}\left(W\partial_rW+Z\partial_{\theta}V\right)
          + \frac{\partial_r\partial_{\theta}W}{2}\,, \\[2ex]
R_{t\theta t\theta} &= \frac{\partial_{r}V\partial_{r}Y}{4X}
          - \frac{\partial_{\theta}V\partial_{\theta}Y}{4Y}
          - \frac{Z(\partial_{\theta}V)^2}{4\mathcal{R}}  + \frac{\partial_{\theta}W}{4\mathcal{R}}\left(V\partial_{\theta}W-W\partial_{\theta}V\right)
          + \frac{\partial_{\theta}\partial_{\theta}V}{2}\,, \\[2ex]
R_{t\theta r\phi} &= \frac{\partial_{\theta}Z}{4\mathcal{R}}\left(W\partial_{r}V-V\partial_{r}W\right)
          - \frac{\partial_{\theta}W}{4\mathcal{R}}\left(W\partial_{r}W+Z\partial_{r}V\right) \\
         &\quad - \frac{\partial_{r}W\partial_{\theta}X}{4X}
          - \frac{\partial_{\theta}W\partial_{r}Y}{4Y}
          + \frac{\partial_{r}\partial_{\theta}W}{2}\,, \\[2ex]
R_{t\theta\theta\phi} &= \frac{\partial_{\theta}Z}{4\mathcal{R}}\left(W\partial_{\theta}V-V\partial_{\theta}W\right)
          - \frac{\partial_{\theta}W}{4\mathcal{R}}\left(W\partial_{\theta}W+Z\partial_{\theta}V\right) \\
         &\quad - \frac{\partial_{r}W\partial_{r}Y}{4X}
          - \frac{\partial_{\theta}W\partial_{\theta}Y}{4Y}
          + \frac{\partial_{\theta}\partial_{\theta}W}{2}\,, \\[2ex]
R_{t\phi t\phi} &= \frac{\partial_{r}V\partial_{r}Z}{4X}
          + \frac{\partial_{\theta}V\partial_{\theta}Z}{4Y}
          + \frac{(\partial_{r}W)^2}{4X}
          + \frac{(\partial_{\theta}W)^2}{4Y}\,, \\[2ex]
R_{t\phi r\theta} &= \frac{\partial_{r}V}{4\mathcal{R}}\left(W\partial_{\theta}Z-Z\partial_{\theta}W\right)
          + \frac{\partial_{r}W}{4\mathcal{R}}\left(Z\partial_{\theta}V-V\partial_{\theta}Z\right) \\
         &\quad + \frac{\partial_{r}Z}{4\mathcal{R}}\left(V\partial_{\theta}W-W\partial_{\theta}V\right)\,, \\[2ex]
R_{r\theta r\theta} &= \frac{(\partial_{\theta}X)^2}{4X}
          + \frac{(\partial_{r}Y)^2}{4Y}
          + \frac{\partial_{r}X\partial_{r}Y}{4X}
          + \frac{\partial_{\theta}X\partial_{\theta}Y}{4Y} - \frac{\partial_{\theta}\partial_{\theta}X}{2}
          - \frac{\partial_{r}\partial_{r}Y}{2}\,, \\[2ex]
R_{r\phi r\phi} &= \frac{\partial_{r}X\partial_{r}Z}{4X}
          - \frac{\partial_{\theta}X\partial_{\theta}Z}{4Y}
          + \frac{V(\partial_rZ)^2}{4\mathcal{R}}
          + \frac{W\partial_{r}W\partial_{r}Z}{2\mathcal{R}}  - \frac{Z(\partial_rW)^2}{4\mathcal{R}}
          - \frac{\partial_r\partial_rZ}{2}\,, \\[2ex]
R_{r\phi\theta\phi} &= \frac{\partial_{r}W}{4\mathcal{R}}\left(W\partial_{\theta}Z-Z\partial_{\theta}W\right)
          + \frac{\partial_{r}Z}{4\mathcal{R}}\left(W\partial_{\theta}W+V\partial_{\theta}Z\right) \\
         &\quad + \frac{\partial_{\theta}X\partial_{r}Z}{4X}
          + \frac{\partial_{r}Y\partial_{\theta}Z}{4Y}
          - \frac{\partial_{\theta}\partial_{\theta}Z}{2}\,, \\[2ex]
R_{\theta\phi\theta\phi} &= \frac{\partial_{\theta}Y\partial_{\theta}Z}{4Y}
          - \frac{\partial_{r}Y\partial_{r}Z}{4X}
          + \frac{V(\partial_{\theta}Z)^2}{4\mathcal{R}}
          + \frac{W\partial_{\theta}W\partial_{\theta}Z}{2\mathcal{R}}  - \frac{Z(\partial_{\theta}W)^2}{4\mathcal{R}}
          - \frac{\partial_{\theta}\partial_{\theta}Z}{2}\,.
\end{align*}

\section{Ricci tensor}

We have the following expressions for the non-zero components of the Ricci tensor
\begin{align*}
R_{tt} &= \frac{\partial_{r}W}{2X\mathcal{R}}(V\partial_{r}W-W\partial_{r}V)
        + \frac{\partial_{\theta}W}{2Y\mathcal{R}}(V\partial_{\theta}W-W\partial_{\theta}V) \\
       &\quad + \frac{\partial_{r}V}{2X\mathcal{R}}(V\partial_{r}Z-Z\partial_{r}V)
        + \frac{\partial_{\theta}V}{2Y\mathcal{R}}(V\partial_{\theta}Z-Z\partial_{\theta}V) \\
       &\quad + \frac{\partial_{r}V\partial_{r}Y}{4XY}
        + \frac{\partial_{\theta}V\partial_{\theta}X}{4XY}
        - \frac{\partial_rV\partial_rX}{4X^2}
        - \frac{\partial_{\theta}V\partial_{\theta}Y}{4Y^2} + \frac{\partial_{r}\partial_{r}V}{2X}
        + \frac{\partial_{\theta}\partial_{\theta}V}{2Y}\,, \\[2ex]
R_{t\phi} &= \frac{\partial_rW}{4X\mathcal{R}}(Z\partial_rV+V\partial_rZ)
        + \frac{\partial_{\theta}W}{4Y\mathcal{R}}(Z\partial_{\theta}V+V\partial_{\theta}Z) \\
       &\quad + \frac{\partial_rW\partial_rX}{4X^2}
        + \frac{\partial_{\theta}W\partial_{\theta}Y}{4Y^2}
        - \frac{\partial_rW\partial_rY}{4XY}
        - \frac{\partial_{\theta}W\partial_{\theta}X}{4XY} \\
       &\quad - \frac{W}{2\mathcal{R}}\left(\frac{\partial_rV\partial_rZ}{X}
        + \frac{\partial_{\theta}V\partial_{\theta}Z}{Y}\right)
        - \frac{\partial_{r}\partial_{r}W}{2X}
        - \frac{\partial_{\theta}\partial_{\theta}W}{2Y}\,, \\[2ex]
R_{rr} &= \frac{W\partial_rW}{\mathcal{R}^2}(Z\partial_rV+V\partial_rZ)
        + \frac{Z^2(\partial_rV)^2}{4\mathcal{R}^2}
        - \frac{W^2\partial_rV\partial_rZ}{2\mathcal{R}^2}
        + \frac{V^2(\partial_rZ)^2}{4\mathcal{R}^2} \\
       &\quad + \frac{(\partial_rW)^2}{2\mathcal{R}}
        + \frac{(\partial_{\theta}X)^2}{4XY}
        + \frac{\partial_rX\partial_rY}{4XY}
        + \frac{(\partial_rY)^2}{4Y^2}
        + \frac{\partial_{\theta}X\partial_{\theta}Y}{4Y^2} \\
       &\quad + \frac{V}{4\mathcal{R}}\left(\frac{\partial_rX\partial_rZ}{X}
        - \frac{\partial_{\theta}X\partial_{\theta}Z}{Y}\right)
        + \frac{W}{2\mathcal{R}}\left(\frac{\partial_rX\partial_rW}{X}
        - \frac{\partial_{\theta}X\partial_{\theta}W}{Y}\right) \\
       &\quad + \frac{Z}{4\mathcal{R}}\left(\frac{\partial_rV\partial_rX}{X}
        - \frac{\partial_{\theta}V\partial_{\theta}X}{Y}\right)
        - \frac{V\partial_{r}\partial_{r}Z}{2\mathcal{R}}
        - \frac{Z\partial_{r}\partial_{\theta}V}{2\mathcal{R}} \\
       &\quad - \frac{W\partial_{r}\partial_{\theta}W}{\mathcal{R}}
        - \frac{\partial_{r}\partial_{r}Y}{2Y}
        - \frac{\partial_{\theta}\partial_{\theta}X}{2Y}\,, \\[2ex]
R_{r\theta} &= \frac{V^2\partial_rZ\partial_{\theta}Z}{4\mathcal{R}^2}
        + \frac{Z^2\partial_rV\partial_{\theta}V}{4\mathcal{R}^2}
        - \frac{W^2}{4\mathcal{R}^2}\left(\partial_rV\partial_{\theta}Z
        + \partial_{\theta}V\partial_rZ\right)
        + \frac{\overline{\mathcal{R}}\,\partial_rW\partial_{\theta}W}{2\mathcal{R}^2} \\
       &\quad + \frac{W}{2\mathcal{R}}\left(\frac{\partial_{\theta}X\partial_rW}{X}
        + \frac{\partial_{\theta}W\partial_rY}{Y}\right)  + \frac{V\partial_rZ+Z\partial_rV}{2\mathcal{R}}
        \left(\frac{W\partial_{\theta}W}{\mathcal{R}}+\frac{\partial_{\theta}X}{2X}\right) \\
       &\quad + \frac{Z\partial_{\theta}V+V\partial_{\theta}Z}{2\mathcal{R}}
        \left(\frac{W\partial_{r}W}{\mathcal{R}}+\frac{\partial_{r}Y}{2Y}\right)  - \frac{V\partial_{r}\partial_{\theta}Z}{2\mathcal{R}}
        - \frac{Z\partial_{r}\partial_{\theta}V}{2\mathcal{R}}
        - \frac{W\partial_{r}\partial_{\theta}W}{\mathcal{R}}\,, \\[2ex]
R_{\theta\theta} &= \frac{W\partial_{\theta}W}{\mathcal{R}^2}(Z\partial_{\theta}V+V\partial_{\theta}Z)
        + \frac{Z^2(\partial_{\theta}V)^2}{4\mathcal{R}^2}
        - \frac{W^2\partial_{\theta}V\partial_{\theta}Z}{2\mathcal{R}^2}
        + \frac{V^2(\partial_{\theta}Z)^2}{4\mathcal{R}^2} \\
       &\quad + \frac{\overline{\mathcal{R}}\,(\partial_{\theta}W)^2}{2\mathcal{R}^2}
        + \frac{(\partial_{\theta}X)^2}{4X^2}
        + \frac{\partial_rX\partial_rY}{4X^2}
        + \frac{(\partial_rY)^2}{4XY}
        + \frac{\partial_{\theta}X\partial_{\theta}Y}{4XY} \\
       &\quad + \frac{V}{4\mathcal{R}}\left(\frac{\partial_{\theta}Y\partial_{\theta}Z}{Y}
        - \frac{\partial_rY\partial_rZ}{X}\right)
        + \frac{W}{2\mathcal{R}}\left(\frac{\partial_{\theta}Y\partial_{\theta}W}{Y}
        - \frac{\partial_rY\partial_rW}{X}\right) \\
       &\quad + \frac{Z}{4\mathcal{R}}\left(\frac{\partial_{\theta}V\partial_{\theta}Y}{Y}
        - \frac{\partial_rV\partial_rY}{X}\right)
        - \frac{V\partial_{\theta}\partial_{\theta}Z}{2\mathcal{R}}
        - \frac{Z\partial_{\theta}\partial_{\theta}V}{2\mathcal{R}} \\
       &\quad - \frac{W\partial_{\theta}\partial_{\theta}W}{\mathcal{R}}
        - \frac{\partial_{\theta}\partial_{\theta}X}{2X}
        - \frac{\partial_{r}\partial_{\theta}Y}{2X}\,, \\[2ex]
R_{\phi\phi} &= \frac{\partial_{r}W}{2X\mathcal{R}}(W\partial_{r}Z-Z\partial_{r}W)
        + \frac{\partial_{\theta}W}{2Y\mathcal{R}}(W\partial_{\theta}Z-Z\partial_{\theta}W) \\
       &\quad + \frac{\partial_{r}Z}{2X\mathcal{R}}(V\partial_{r}Z-Z\partial_{r}V)
        + \frac{\partial_{\theta}Z}{2Y\mathcal{R}}(V\partial_{\theta}Z-Z\partial_{\theta}V) \\
       &\quad - \frac{\partial_{r}Y\partial_{r}Z}{4XY}
        - \frac{\partial_{\theta}X\partial_{\theta}Z}{4XY}
        + \frac{\partial_rX\partial_rZ}{4X^2}
        + \frac{\partial_{\theta}Y\partial_{\theta}Z}{4Y^2}  - \frac{\partial_{r}\partial_{r}Z}{2X}
        - \frac{\partial_{\theta}\partial_{\theta}Z}{2Y}
\end{align*}
where $\overline{\mathcal{R}}=W^2-VZ$.

\section{Series expansions and auxiliary quantities}
\label{app:series}
As established in the main text, under the regime of the supplementary condition employed to solve the equations, it is valid to retain terms only up to linear order in the spin. Nevertheless, terms of order $s^2$ may appear in these appendices as a consequence of the full series expansions, and they have been included for the sake of completeness. We do \textbf{not} claim that these higher-order terms represent the complete physical description at second order in the spin.

Expression for $p_t$ and $p_\phi$ in terms of the energy and the angular momentum are
\begin{align*}
    \mathcal{E}_t &= - 1 - \frac{Ms(am + s)}{m^2} u^3 - \frac{3qs^2}{2m^2} u^5 - \frac{Mqs(7am + 5s)}{2m^2} u^6,\\[2ex]
    \mathcal{L}_t &= \frac{M s}{m} u^3 + \frac{3 q s}{2 m} u^5 + \frac{9 M q s}{2 m} u^6,\\[2ex]
    \mathcal{E}_{\phi} &=-\frac{s}{m} + \frac{s(a^2 M + 2q)}{m} u^3 + \frac{33 M q s}{4m} u^4 - \frac{5 q^2 s}{4m} u^6,\\[2ex]
    \mathcal{L}_{\phi} &= 1 + \frac{Ms(s-am )}{m^2} u^3 + \frac{3qs^2}{2m^2} u^5 + \frac{Mqs(5s-7am)}{2m^2} u^6.
\end{align*}

The coefficients of the effective potential defined in (\ref{v}) are given by
\begin{align*}
     A &= -2\left(1 - 2Mu + \frac{J^2}{M^2}u^2 + qu^3 - \frac{1}{2} Mqu^4 + \frac{5}{4} q^2 u^6\right),\\[2ex]
     B &= u^2 - 2Mu^3 + 2qu^5 + \frac{1}{2} Mqu^6 + \frac{11}{4} q^2 u^8 - \hat{s}\left(2Ju^5 + 17Jqu^8\right) + \hat{s}^2\left(2Mu^5 - 5M^2u^6 + 3qu^7\right),\\[2ex]
     C &= 2\left[ 2J u^3 + \frac{7}{2} J q u^6 
    + \hat{s} \left( -u^2 + 3M u^3 + 2\frac{J^2}{M} u^5 + \frac{3}{2} q u^5 + \frac{33}{4} M q u^6 \right) 
    + \hat{s}^2 \left( J u^5 + 5 J M u^6\right)\right],\\[2ex]
     D &= -1 - \frac{J^2}{M^2} u^2 - 2\frac{J^2}{M} u^3 + \frac{3}{2} M q u^4 - \frac{3}{4} q^2 u^6 
    - \hat{s} \left( 6 J u^3 + 2\frac{J^3}{M^2} u^5 + 6 J q u^6 \right) \\
    &\quad- \hat{s}^2\left( -u^2 + 4 M u^3 + 4\frac{J^2}{M} u^5 + 5 q u^5 + 5 J^2 u^6 + 13 M q u^6 \right).
\end{align*}
The following expressions correspond to the derivatives of the preceding coefficients
\begin{align*}
    A'&=-2\left(- 2M + 2\frac{J^2}{M^2}u + 3qu^2 - 2Mqu^3 + \frac{15}{2} q^2 u^5\right),\\[2ex]
    A''&=-2\left(2\frac{J^2}{M^2} + 6qu - 6Mqu^2 + \frac{75}{2} q^2 u^4\right),\\[2ex]
    B'&= 2u - 6Mu^2 + 10qu^4 + 3Mq u^5 + 22q^2 u^7 - \hat{s} \left( 2Ju^4(5 + 68qu^3) \right) + \hat{s}^2u^4 \left( 10M - 30M^2u + 21qu^2 \right),\\[2ex]
    B''&= 2 - 12Mu + 40qu^3 + 15Mq u^4 + 154q^2 u^6 - \hat{s} \left( 8Ju^3(5 + 119qu^3) \right) + \hat{s}^2 u^3 \left( 40M - 150M^2u + 126qu^2 \right),\\[2ex]
    C'&= 2\left[ 3Ju^2(2 + 7qu^3) + \hat{s} \left( -2u + 9Mu^2 + \frac{15}{2}qu^4 + 10\frac{J^2}{M}u^4 + \frac{99}{2}Mqu^5 \right) + \hat{s}^2 \left( 5Ju^4(1 + 6Mu) \right)\right],\\[2ex]
    C''&=2\left[ 3Ju(4 + 35qu^3) + \hat{s} \left( -2 + 18Mu + 30qu^3 + 40\frac{J^2}{M}u^3 + \frac{495}{2} Mqu^4 \right) + \hat{s}^2 \left( 10Ju^3(2 + 15Mu) \right)\right],\\[2ex]
    D'&= -2\frac{J^2}{M^2} u - 6\frac{J^2}{M} u^2 + 6Mq u^3 - \frac{9}{2} q^2 u^5 - \hat{s} \left( 18 J u^2 + 10\frac{J^3}{M^2} u^4 + 36 J q u^5 \right) \\
    &\quad- \hat{s}^2 \left( -2u + 12 M u^2 + 20\frac{J^2}{M} u^4 + 25 q u^4 + 30 J^2 u^5 + 78 M q u^5 \right),\\[2ex]
    D''&= -2\frac{J^2}{M^2} - 12\frac{J^2}{M} u + 18Mq u^2 - \frac{45}{2} q^2 u^4 - \hat{s} \left( 36 J u + 40\frac{J^3}{M^2} u^3 + 180 J q u^4 \right) \\
    &\quad- \hat{s}^2 \left( -2 + 24 M u + 80\frac{J^2}{M} u^3 + 100 q u^3 + 150 J^2 u^4 + 390 M q u^4 \right).
\end{align*}

The functions $F$ and all their distinct combinations forming the auxiliary quantities are given by
\begin{align*}
    F_1 &= - 4 u + 16 M u^{2} - 16 M^{2} u^{3} + \left(4 M a^{2} + 20 M a \hat{s} - 20 M \hat{s}^2 - 18 q\right) u^{4} + \left(- 32 M^{2} a \hat{s}^2 + 92 M^{2} \hat{s}^2 + 24 M q\right) u^{5} \\
    &\quad - \left(100 M^{3} \hat{s}^2 - 12 M a^{3}\hat{s} + 12 M a^{2}\hat{s}^2 + 42 q \hat{s}^2\right) u^{6}  + \left(40 M^{2} a^{2}\hat{s}^2 + 280 M a q \hat{s} + 64 M q \hat{s}^2 - 42 q^{2}\right) u^{7} - 24 q^{2}\hat{s}^2 u^{9},\\[2ex]
    F_2 &= 8 \hat{s} u - \big(24 M a + 44 M \hat{s}\big) u^{2}
    + \left(32 M^{2} a + 48 M^{2} \hat{s}\right) u^{3}  - \left(8 M a^{3} + 52 M a^{2} \hat{s} + 20 M a \hat{s}^{2} + 34 q \hat{s}\right) u^{4} \\
    &\quad + \big(64 M^{2} a^{2} \hat{s} - 88 M^{2} a \hat{s}^{2} - 84 M a q - 146 M q \hat{s}\big) u^{5}  + \left(200 M^{3} a \hat{s}^{2} - 24 M a^{4} \hat{s} - 12 M a^{3} \hat{s}^{2}\right) u^{6} \\
    &\quad - \left(80 M^{2} a^{3} \hat{s}^{2} + 456 M a q \hat{s}^{2} + 104 q^{2} \hat{s}\right) u^{7},\\[2ex]
    F_3 &= 4 M - 4 \hat{s}^{2} u + \left(8 M a^{2} + 36 M a \hat{s} + 28 M \hat{s}^{2} - 6 q\right) u^{2}  - \left(16 M^{2} a^{2} + 48 M^{2} a \hat{s} + 32 M^{2} \hat{s}^{2} + 8 M q\right) u^{3} \\
    &\quad + \left(4 M a^{4} + 32 M a^{3} \hat{s} + 48 M a^{2} \hat{s}^{2} + 52 q \hat{s}^{2}\right) u^{4} + \big(- 32 M^{2} a^{3} \hat{s} - 4 M^{2} a^{2} \hat{s}^{2} + 72 M a q \hat{s} + 74 M q \hat{s}^{2} - 6 q^{2}\big) u^{5} \\
    &\quad + \left(- 100 M^{3} a^{2} \hat{s}^{2} + 12 M a^{5} \hat{s} + 24 M a^{4} \hat{s}^{2}\right) u^{6} + \left(40 M^{2} a^{4} \hat{s}^{2} - 14 q^{2} \hat{s}^{2}\right) u^{7},\\[2ex]
    F_4 &= - \left(4 M a + 2 M \hat{s}\right) u^{4} - \left(12 M a^{2} \hat{s} - 6 M a \hat{s}^{2} + 21 q \hat{s}\right) u^{6} - \left(40 M^{2} a \hat{s}^{2} + 12 M a q + 34 M q \hat{s}\right) u^{7} \\
    &\quad + \left(84 M a q \hat{s}^{2} - 60 q^{2} \hat{s}\right) u^{9},\\[2ex]
    F_5 &= 2 u - 6 M u^{2} - \big(4 M a^{2} + 16 M a \hat{s} - 8 M \hat{s}^{2} - 10 q\big) u^{4} - \left(30 M^{2} \hat{s}^{2} - 6 M q\right) u^{5} - \left(12 M a^{3} \hat{s} + 6 M a^{2} \hat{s}^{2}\right) u^{6} \\
    &\quad - \left(40 M^{2} a^{2} \hat{s}^{2} + 136 M a q \hat{s} + 18 M q \hat{s}^{2} - 19 q^{2}\right) u^{7},\\[2ex]
    F_6 &= - 4 \hat{s} u + \left(12 M a + 18 M \hat{s}\right) u^{2} + \left(4 M a^{3} + 30 M a^{2} \hat{s} + 18 M a \hat{s}^{2} + 15 q \hat{s}\right) u^{4} + \left(60 M^{2} a \hat{s}^{2} + 42 M a q + 93 M q \hat{s}\right) u^{5} \\
    &\quad + \left(12 M a^{4} \hat{s} + 18 M a^{3} \hat{s}^{2}\right) u^{6} + \left(40 M^{2} a^{3} \hat{s}^{2} + 240 M a q \hat{s}^{2} + 42 q^{2} \hat{s}\right) u^{7},\\[2ex]
    \Delta_1 &= 64 M u - 256 M^{2} u^{2}  + \left(256 M^{3} + 128 M a^{2} + 192 M a \hat{s} - 96 q\right) u^{3} \\
    &\quad - \left(256 M^{2} a^{2} + 768 M^{2} a \hat{s} - 464 M^{2} \hat{s}^{2} - 544 M q\right) u^{4} + \left(768 M^{3} a \hat{s} - 1856 M^{3} \hat{s}^{2} + 64 M a^{4} + 384 M a^{3} \hat{s}\right) u^{5} \\
    &\quad + \left(1856 M^{4} \hat{s}^{2} - 768 M^{2} a^{3} \hat{s} + 928 M^{2} a^{2} \hat{s}^{2} + 4512 M a q \hat{s} + 1104 M q \hat{s}^{2} - 528 q^{2}\right) u^{6} \\
    &\quad + \left(- 1856 M^{3} a^{2} \hat{s}^{2} + 192 M a^{5} \hat{s}\right) u^{7} + \left(464 M^{2} a^{4} \hat{s}^{2} + 1332 q^{2} \hat{s}^{2}\right) u^{8},\\[2ex]
    \Delta_2 &= 4 u^{2} - 24 M u^{3} + 36 M^{2} u^{4} - \left(16 M a^{2} + 48 M a \hat{s} - 40 M \hat{s}^{2} - 40 q\right) u^{5}+ \left(96 M^{2} a \hat{s} - 252 M^{2} \hat{s}^{2} - 96 M q\right) u^{6} \\
    &\quad + \left(360 M^{3} \hat{s}^{2} - 48 M a^{3} \hat{s} + 24 M a^{2} \hat{s}^{2} + 84 q \hat{s}^{2}\right) u^{7} - \left(172 M^{2} a^{2} \hat{s}^{2} + 1128 M a q \hat{s} + 184 M q \hat{s}^{2} - 176 q^{2}\right) u^{8} \\
    &\quad- 75 q^{2} \hat{s}^{2} u^{10},\\[2ex]
    \Pi &= - 16 u^{2} + 112 M u^{3} - 256 M^{2} u^{4} + \left(192 M^{3} + 48 M a^{2} + 176 M a \hat{s} - 160 M \hat{s}^{2} - 152 q\right) u^{5} \\
    &\quad + \left(- 80 M^{2} a^{2} - 656 M^{2} a \hat{s} + 1192 M^{2} \hat{s}^{2} + 584 M q\right) u^{6} \\
    &\quad + \left(640 M^{3} a \hat{s} - 2816 M^{3} \hat{s}^{2} + 144 M a^{3} \hat{s} - 96 M a^{2} \hat{s}^{2} - 336 q \hat{s}^{2}\right) u^{7} \\
    &\quad + \left(2160 M^{4} \hat{s}^{2} - 304 M^{2} a^{3} \hat{s} + 584 M^{2} a^{2} \hat{s}^{2} + 3728 M a q \hat{s} + 936 M q \hat{s}^{2} - 680 q^{2}\right) u^{8} - 720 M^{3} a^{2} \hat{s}^{2} u^{9} \\
    &\quad - \left(144 M^{2} a^{4} \hat{s}^{2} + 702 q^{2} \hat{s}^{2}\right) u^{10},\\[2ex]
    F_7 &= 128 M u - \left(1280 M^{2} + 128 a^{2}\right) u^{2} + \left(3584 M^{3} + 512 M a^{2} - 384 M a \hat{s} - 192 q\right) u^{3} \\
    &\quad + \left(- 3072 M^{4} - 2432 M^{2} a^{2} - 3712 M^{2} a \hat{s} + 2624 M^{2} \hat{s}^{2} + 5952 M q\right) u^{4} \\
    &\quad + \left(3328 M^{3} a^{2} + 17152 M^{3} a \hat{s} - 25472 M^{3} \hat{s}^{2} + 384 M a^{4} - 512 M a^{3} \hat{s} - 1280 M a^{2} \hat{s}^{2}\right) u^{5} \\
    &\quad - \big(17408 M^{4} a \hat{s} - 68992 M^{4} \hat{s}^{2} + 1024 M^{2} a^{4} + 8704 M^{2} a^{3} \hat{s} - 512 M^{2} a^{2} \hat{s}^{2} + 18048 M a q \hat{s} + 2688 M q \hat{s}^{2} + 6336 q^{2}\big) u^{6} \\
    &\quad + \big(- 57600 M^{5} \hat{s}^{2} + 17920 M^{3} a^{3} \hat{s} - 16640 M^{3} a^{2} \hat{s}^{2} + 1152 M a^{5} \hat{s} - 768 M a^{4} \hat{s}^{2}\big) u^{7} \\
    &\quad + \big(33408 M^{4} a^{2} \hat{s}^{2} - 5504 M^{2} a^{5} \hat{s} - 4928 M^{2} a^{4} \hat{s}^{2} - 42192 q^{2} \hat{s}^{2}\big) u^{8}  + 8064 M^{3} a^{4} \hat{s}^{2} u^{9} - 6912 M^{2} a^{6} \hat{s}^{2} u^{10},\\[2ex]
    F_8 &= \left(96 M a + 48 M \hat{s}\right) u^{2} - \left(64 M^{2} a + 32 M^{2} \hat{s}\right) u^{3} + \left(- 32 M a^{3} + 464 M a^{2} \hat{s} - 240 M a \hat{s}^{2} + 840 q \hat{s}\right) u^{4} \\
    &\quad + \big(- 384 M^{2} a^{2} \hat{s} + 2112 M^{2} a \hat{s}^{2} + 288 M a q + 480 M q \hat{s}\big) u^{5}  + \left(- 1600 M^{3} a \hat{s}^{2} - 96 M a^{4} \hat{s} + 48 M a^{3} \hat{s}^{2}\right) u^{6} \\
    &\quad + \left(- 320 M^{2} a^{3} \hat{s}^{2} - 6144 M a q \hat{s}^{2} + 3168 q^{2} \hat{s}\right) u^{7},\\[2ex]
    \Delta_3 &= -\left(1536 M a + 768 M \hat{s}\right) u^{4} + \left(12288 M^{2} a + 6144 M^{2} \hat{s}\right) u^{5} \\
    &\quad - \left(36864 M^{3} a + 18432 M^{3} \hat{s} + 7680 M a^{2} \hat{s} - 3840 M a \hat{s}^{2} + 13440 q \hat{s}\right) u^{6} \\
    &\quad + \big(49152 M^{4} a + 24576 M^{4} \hat{s} + 3072 M^{2} a^{3} + 78336 M^{2} a^{2} \hat{s} - 69120 M^{2} a \hat{s}^{2} - 19968 M a q + 81408 M q \hat{s}\big) u^{7} \\
    &\quad + \left(- 24576 M^{5} a - 12288 M^{5} \hat{s} - 12288 M^{3} a^{3} - 276480 M^{3} a^{2} \hat{s} + 427008 M^{3} a \hat{s}^{2}\right) u^{8} \\
    &\quad + \big(12288 M^{4} a^{3} + 411648 M^{4} a^{2} \hat{s} - 1201152 M^{4} a \hat{s}^{2} + 24576 M^{2} a^{4} \hat{s} + 64512 M^{2} a^{3} \hat{s}^{2} + 243456 M a q \hat{s}^{2} \\
    &\qquad - 182400 q^{2} \hat{s}\big) u^{9} \\
    &\quad + \big(- 221184 M^{5} a^{2} \hat{s} + 1585152 M^{5} a \hat{s}^{2} - 1536 M^{3} a^{5} - 114432 M^{3} a^{4} \hat{s} - 319488 M^{3} a^{3} \hat{s}^{2}\big) u^{10} \\
    &\quad + \big(- 798720 M^{6} a \hat{s}^{2} + 122880 M^{4} a^{4} \hat{s} + 445440 M^{4} a^{3} \hat{s}^{2} + 46080 M^{2} a^{5} \hat{s}^{2}\big) u^{11} \\
    &\quad - \big(144384 M^{5} a^{3} \hat{s}^{2} + 16896 M^{3} a^{6} \hat{s} + 298752 M^{3} a^{5} \hat{s}^{2}\big) u^{12} \\
    &\quad + \left(319488 M^{4} a^{5} \hat{s}^{2}\right) u^{13}  - 59904 M^{3} a^{7} \hat{s}^{2} u^{14},\\[2ex]
    \Delta_4 &= - 2048 M u^{3} + 34816 M^{2} u^{4} + \left(- 233472 M^{3} + 6144 M a^{2} + 6144 M a \hat{s} - 3072 q\right) u^{5} \\
    &\quad + \left(802816 M^{4} - 30720 M^{2} a^{2} + 38912 M^{2} a \hat{s} - 62464 M^{2} \hat{s}^{2} - 44032 M q\right) u^{6} \\
    &\quad + \left(- 1507328 M^{5} - 20480 M^{3} a^{2} - 876544 M^{3} a \hat{s} + 1064960 M^{3} \hat{s}^{2} + 30720 M a^{3} \hat{s}\right) u^{7} \\
    &\quad + \big(1474560 M^{6} + 385024 M^{4} a^{2} + 4513792 M^{4} a \hat{s} - 7116800 M^{4} \hat{s}^{2} - 12288 M^{2} a^{4} - 227328 M^{2} a^{3} \hat{s} \\
    &\qquad + 125952 M^{2} a^{2} \hat{s}^{2} + 448512 M a q \hat{s} - 92160 M q \hat{s}^{2} - 24576 q^{2}\big) u^{8} \\
    &\quad + \big(- 589824 M^{7} - 802816 M^{5} a^{2} - 10600448 M^{5} a \hat{s} + 24358912 M^{5} \hat{s}^{2} + 18432 M^{3} a^{4} + 32768 M^{3} a^{3} \hat{s} \\
    &\qquad - 1492992 M^{3} a^{2} \hat{s}^{2}\big) u^{9} \\
    &\quad + \big(524288 M^{6} a^{2} + 11993088 M^{6} a \hat{s} - 45539328 M^{6} \hat{s}^{2} + 83968 M^{4} a^{4} + 2887680 M^{4} a^{3} \hat{s} \\
    &\qquad + 6178816 M^{4} a^{2} \hat{s}^{2} - 98304 M^{2} a^{5} \hat{s} - 202752 M^{2} a^{4} \hat{s}^{2} + 481536 q^{2} \hat{s}^{2}\big) u^{10} \\
    &\quad + \big(- 5308416 M^{7} a \hat{s} + 44384256 M^{7} \hat{s}^{2} - 151552 M^{5} a^{4} - 7168000 M^{5} a^{3} \hat{s} - 10924032 M^{5} a^{2} \hat{s}^{2} \\
    &\qquad + 6144 M^{3} a^{6} + 202752 M^{3} a^{5} \hat{s} - 528384 M^{3} a^{4} \hat{s}^{2}\big) u^{11} \\
    &\quad + \big(- 17694720 M^{8} \hat{s}^{2} + 5144576 M^{6} a^{3} \hat{s} + 7098368 M^{6} a^{2} \hat{s}^{2} + 14336 M^{4} a^{6} + 813056 M^{4} a^{5} \hat{s} \\
    &\qquad + 8369152 M^{4} a^{4} \hat{s}^{2} - 184320 M^{2} a^{6} \hat{s}^{2}\big) u^{12} \\
    &\quad + \big(- 122880 M^{7} a^{2} \hat{s}^{2} - 1617920 M^{5} a^{5} \hat{s} - 19587072 M^{5} a^{4} \hat{s}^{2} + 67584 M^{3} a^{7} \hat{s} + 522240 M^{3} a^{6} \hat{s}^{2}\big) u^{13} \\
    &\quad + \big(12990464 M^{6} a^{4} \hat{s}^{2} + 165888 M^{4} a^{7} \hat{s} + 3204096 M^{4} a^{6} \hat{s}^{2}\big) u^{14} + \left(- 5793792 M^{5} a^{6} \hat{s}^{2} + 239616 M^{3} a^{8} \hat{s}^{2}\right) u^{15} \\
    &\quad + 712704 M^{4} a^{8} \hat{s}^{2} u^{16}.
\end{align*}
Using these definitions and their combinations, one can construct general expressions for $\mathcal{P}_{\text{ISCO}}$, $E^2$, and $L_z^2$, and subsequently truncate the relevant parameters to recover the corresponding formulas for each desired limiting case.

\twocolumngrid
\bibliography{bibtex}

\end{document}